\documentclass[12pt]{article}
\usepackage{amsmath,amsfonts, epsfig}
\usepackage{fullpage}
\usepackage{amssymb}
\usepackage{amscd}
\advance \textheight by \topskip
\makeatletter

\@addtoreset{equation}{section}
 \def\theequation{\thesection.\arabic{equation}}
\makeatother

\def\CC{{\mathbb C}}
\def\RR{{\mathbb R}}

\newcommand{\beqa}{\begin{eqnarray}}
\newcommand{\eeqa}{\end{eqnarray}}
\newcommand{\noi}{\noindent}

\newcommand{\lp}{\left(}
\newcommand{\rp}{\right)}

\newcommand{\tri}{\includegraphics[scale=0.07]{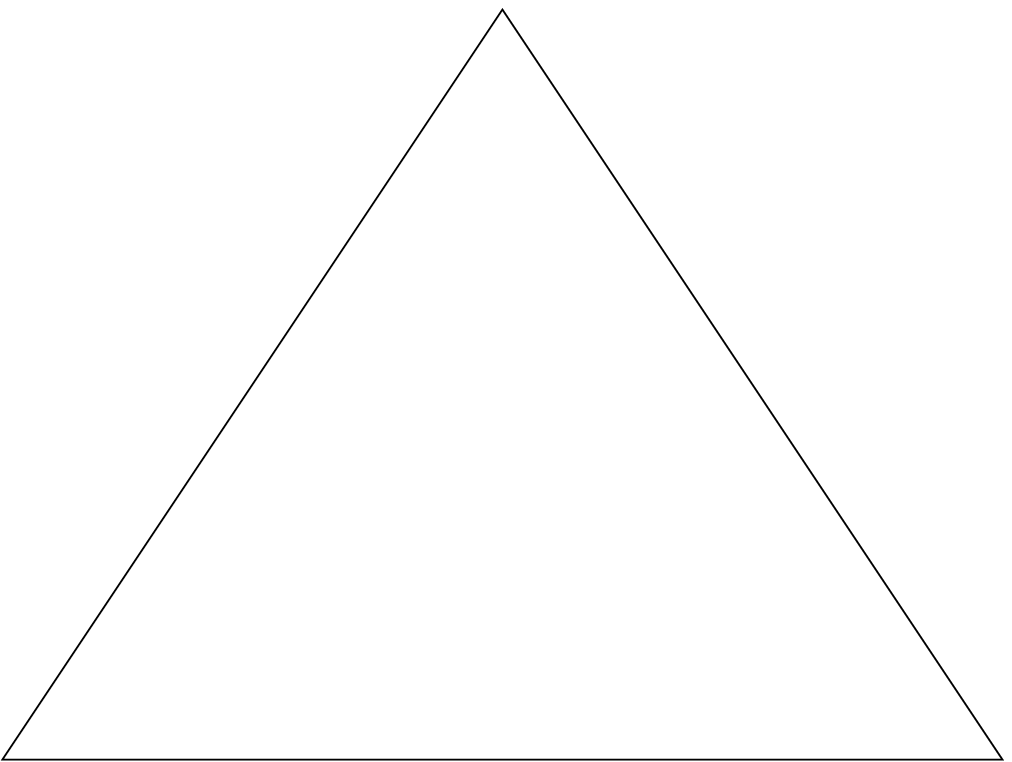}}

\newcommand{\T}{{\mathbb T}}
\newcommand{\unu}{1_{\mathbb T}}

\newcommand{\ba}{\begin{array}}
\newcommand{\ea}{\end{array}}

\newcommand{\fig}{\epsfig}

\def\>{\rangle}
\def\<{\langle}

\begin{document}

\title{
\begin{flushright}
\end{flushright}
{\bf Algebraic structures in quantum gravity}
\author{ 
{\sf   Adrian Tanasa${}^{}$\thanks{e-mail:
adrian.tanasa@ens-lyon.org}
}\\
{\small ${}^{a}${\it Centre de Physique Th\'eorique, CNRS, UMR 7644,}} \\
{\small {\it \'Ecole Polytechnique, 91128 Palaiseau, France}}  \\ 
{\small ${}^{b}${\it Institutul de Fizic\u a \c si Inginerie Nuclear\u a Horia Hulubei,}} \\
{\small {\it P. O. Box MG-6, 077125 M\u agurele, Rom\^ania}}\\
\\
}}
\maketitle
\vskip-1.5cm

\vspace{2truecm}

\begin{abstract}
\noindent
Starting from a recently-introduced algebraic structure on spin foam models, we define a Hopf algebra by dividing with an appropriate quotient. 
The structure, thus defined, naturally allows for a mirror analysis of spin foam models with quantum field theory, from a combinatorial point of view. 
A grafting operator is introduced allowing for the equivalent of a Dyson-Schwinger equation to be written.
Non-trivial examples are explicitly worked out. Finally, the physical significance of the results is discussed.
\end{abstract}


\newpage

\section{Introduction and motivation}
\renewcommand{\theequation}{\thesection.\arabic{equation}}    
\setcounter{equation}{0}

Formulating a renormalizable quantum theory for gravity is perhaps the most important open question of contemporary fundamental physics.  Noncommutative geometry \cite{book-connes} 
can be requested when quantum mechanics and gravity meet at some energy scale \cite{dfr}.
String theory, loop quantum gravity, dynamical triangulations {\it etc.} have made, in the last few decades, different propositions for new physics such that this crucial task of unification can be achieved.

When considering loop quantum gravity (see for example \cite{book}), the historic way to approach the quantification 
 is to use the spin foam (SF) formalism.
Lately it has also been indicated that this formalism can be equivalent to a new type of formulation, the group field theoretical one (see for example \cite{gft}).
 
In this paper, we investigate  some algebraic properties of the SF formulation of loop quantum gravity. The starting point is the algebraic structure of SF models introduced in \cite{fm} (following previous work in \cite{fm-primu}), structure related to the Connes-Kreimer algebra. Note that in commutative or resp. noncommutative quantum field theory (QFT) it was proved that the Connes-Kreimer Hopf algebra of Feynman graphs gives rise to the celebrated BPHZ forest formula (\cite{ck} and resp. \cite{fab,io-kreimer}).
Nevertheless, a key ingredient of renormalizability is the power counting theorem, which tells us which are the primitive divergent graphs to sum on in the definition of the Connes-Kreimer coproduct.

This power counting result for SF models is not known and therefore is not present in the construction proposed in \cite{fm}. Moreover, in commutative QFT the notion of locality is essential for the renormalizability of the model (see for example \cite{book2}). This notion generalizes in the case of noncommutative QFT to the notion of ``Moyality'' (one has non-local counterterms of the same form as the original non-local ones in the initial, noncommutative action, see  for example \cite{rev}  or \cite{beta-GMRT} for details). 

The Connes-Kreimer Hopf algebra underlying renormalization has recently been extended to a more general Hopf algebra, the core Hopf algebra \cite{core},  where the  coproduct  sums over {\it any} subgraph.  This core Hopf algebra was also introduced within the framework of  noncommutative QFT  \cite{io-kreimer}.
This implies that the only Hopf primitives (that is the graphs which have a trivial coproduct) are the $1-$loop graphs.

\medskip

In this paper, we start from the construction proposed in \cite{fm} and we quotient out a Hopf coideal in order to obtain a new algebraic structure whose properties are more naturally interconnected to the algebraic properties one is familiar with in (non)commutative QFT.
This new construction is easily proved to be a Hopf algebra; its  graduation structure will be explained here.
Furthermore, we notice  that  this  algebra
can be interpreted as the {\it core Hopf algebra of SFs}, since in the coproduct (just as in the one introduced in \cite{fm}) one sums over {\it all} sub-SFs.


To further support this idea comes the remark that, when dealing with perturbative gravity, the core Hopf algebra is the pertinent Hopf algebra structure, because the one-loop graphs are the Hopf primitives  (that is the graphs which have a trivial coproduct) \cite{core-suite}.

Let us also emphasize that in a commutative or noncommutative QFT, once one has a Hopf algebra structure, one can define some grafting operator 
$B_+$. In the language of Feynman diagrams of (non)commutative QFT, this corresponds to the  operator of insertions of subgraphs into graphs. To any  primitively divergent graph in a (non)commutative QFT model one can associate such an operator.
Any relevant graph in perturbation theory is then in the image of such an operator $B_+$.
This property is intimately related to the physical principle of locality in commutative QFT \cite{bk} or to the one of ``Moyality'' in noncommutative QFT \cite{io-kreimer}.
One can then write down the combinatorial Dyson-Schwinger equation in a recursive way, as a power series written in terms of these insertion operators $B_+$.  When applying the renormalized Feynman rules to the combinatorial Dyson-Schwinger equations in QFT, one deals with the usual analytic Dyson-Schwinger equations 
\cite{bk, yeats}.

Recently, within the QFT core Hopf algebra setting, the role of the same operator $B_+$ has been thoroughly investigated \cite{core-suite}. Moreover, the structure of Dyson-Schwinger equations in the perturbative quantum field theory of gravity has been recently studied in \cite{k-qg} and it was suggested that gravity, regarded as a probability conserving but perturbatively non-renormalizable theory, is renormalizable after all.

In this paper, we define  an appropriate grafting operator $B_+$ and we perform this type of analysis for SFs in $2$, $3$ and $4D$. We propose a way of adapting all of these notions of (non)commutative QFT for this completely different setting. 
The physical meaning of these results is however related to a possible generalization of the locality (or ``Moyality'') notions mentioned above.  We will argue further on that in the conclusion section of this paper.


\medskip

This paper is structured as follows. In the next section, we recall from \cite{fm} the algebraic construction proposed there. We then define, in the third section,  for $2D$ SFs the Hopf algebra $\T$ obtained from the construction of \cite{fm} by taking some apropriate quotient. The graduation of $\T$ is presented and the grafting operator $B_+$ is defined. We then give a list of the algebraic properties existing in $\T$, properties which are in perfect analogy with the ones existing in (non)commutative QFT. We also explicitly work out some non-trivial examples which illustrate these properties. 
In the following section, the generalizations of these results to $3$ and $4D$ SFs is presented. The last section is dedicated to the conclusions and to a final discussion.

\section{SFs; partitioned SFs and parenthesized weights}
\renewcommand{\theequation}{\thesection.\arabic{equation}}    
\setcounter{equation}{0}



A SF is a combinatorial object which can be seen as the world-surface swept by a spin network. The spin networks are graphs labeled by the representations of some group (edges are labeled by representations and nodes are labeled by intertwiners). This implies that the faces of the SF are labeled by representations, the edges by the intertwiners; the vertices carry the evolution amplitudes.
A SF represents a space-time.

Consider now the following partition function, defined as the sum:
\beqa
Z(s_i, s_f)=\sum_\Gamma N(\Gamma) \sum_{{\rm labels\, on}\,\Gamma}\prod_{f\in\Gamma}{\rm dim}\, j_f\prod_{v\in\Gamma}A_v(j).
\eeqa
We have denoted by $s_i$ and resp. $s_f$ the initial and resp. the final spin networks between which SFs $\Gamma$ interpolate. A face of the SF is denoted by $f$ and the dimension of the group representation $j$ labeling it by dim $j_f$. The function $A_v$ is the vertex amplitude and is associated with any vertex $v$ of the SF. Finally, $N$ is a weight factor depending only on
the SF itself.

We also denote by
\beqa
\omega_\Gamma=\prod_{f\in\Gamma}{\rm dim}\, j_f\prod_{v\in\Gamma}A_v(j)
\eeqa
the weight of the respective SF. It is this weight which encodes the physical content of the SF.

Choosing the set of SFs $\Gamma$, associated factors $N(\Gamma)$, the set of representations and intertwiners as well as the amplitudes $A_v$, defines the respective SF model. For a general review of SFs, the interested reader may refer himself, for example, to \cite{sf}. The EPRL \cite{eprl} and Freidel-Krasnov \cite{fk} models are the current SF models candidate to describe a microscopic structure of space-time and to have a good low energy limit (which contains the known theories).


\medskip

We now follow \cite{fm} to define partitioned SFs and paranthesized SFs.
A sub-SF $\gamma$ of a SF $\Gamma$ is a subset of faces of $\Gamma$, together with any vertices and edges that are boundaries of these faces.

A sub-SF $\gamma_1$ is nested into a sub-SF $\gamma_2$, $\gamma_1\subset\gamma_2$ if the set of faces of $\gamma_1$ is a proper subset of faces of $\gamma_2$.

Two sub-SFs $\gamma_i$ ($i=1,2$) are disjoint sub-SFs, $\gamma_1\cap\gamma_2=\emptyset$, if and only if they have no faces, edges or vertices in common. 

One says that two sub-SFs are not overlapping if the respective sub-SFs are either nested or disjoint. Furthermore, 
 an allowed partition into sub-SFs of a SF is a partition for which any two sub-SFs are not overlapping. 

A partitioned SF is a SF  marked with an allowed partitioned into sub-SFs.
We denote by $\Gamma/\gamma$ a co-SF, that is the SF obtained from shrinking the sub-SF $\gamma$ of the SF $\Gamma$ into a single vertex.

As in \cite{fm}, we will work out in this paper with partitioned SFs, referred however to as SFs.

\begin{figure}
\centerline{\epsfig{figure=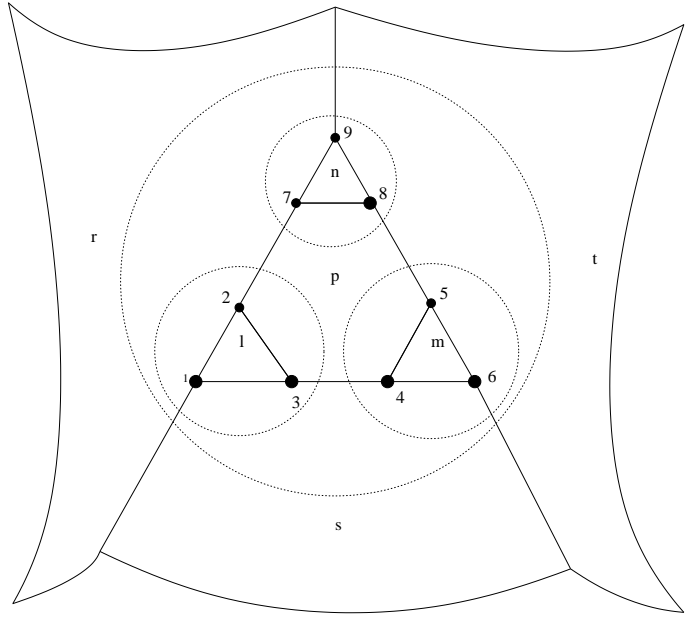,width=9cm} }
\caption{An example of a $2$-dimensional SF $\Gamma$  and the way to represent its parenthesized weight.}
\label{weight}
\end{figure}

The weight of a given SF is represented, as explained in \cite{fm} by a parenthesized weight. For example, for the SF $\Gamma$ of Fig. \ref{weight}
one has
\beqa
\omega_\Gamma&=&\left(\left(\left( \omega_{\gamma_1}\right)\left(\omega_{\gamma_2}\right)\left(\omega_{\gamma_3}\right)\omega_{\Gamma'/\gamma_1\cup\gamma_2\cup\gamma_3}\right)\omega_{\Gamma/\Gamma'}\right),\nonumber\\
&=&\left(\left(\left( d_l A_{v_1}A_{v_2}A_{v_3}\right) \left( d_m A_{v_4}A_{v_5}A_{v_6}\right)\left( d_n A_{v_7}A_{v_8}A_{v_9}\right)d_p\right)d_rd_sd_t\right),
\eeqa
where $\gamma_1$, $\gamma_2$ and resp. $\gamma_3$ are the sub-SFs with faces $l$, $m$ and resp. $n$ and $\Gamma'$ is the sub-SF containing $\gamma_i$ ($i=1,2,3$), see again Fig. \ref{weight}.

In \cite{fm}, on the space of these SFs was defined a coproduct
\beqa
\label{coproductM}
\Delta_M \Gamma= \Gamma \otimes 1_M + 1_M \otimes \Gamma + \sum_{\gamma\subset\Gamma} \gamma\otimes \Gamma/\gamma,
\eeqa 
where $\Gamma$ is some SF and $\gamma$ any of its SFs. Finally, we have denoted by $1_M$ the empty SF.

\section{The core Hopf algebra  and the grafting operator -  definition and mirror analysis with QFT}

In this section we focus on the $2D$ SFs. 
In \cite{fm}, when applying the coproduct $\Delta_M$ one has  (see Example $1$ of \cite{fm})
\beqa
\Delta'_M \lp \begin{array}{c}\mbox{\epsfig{file=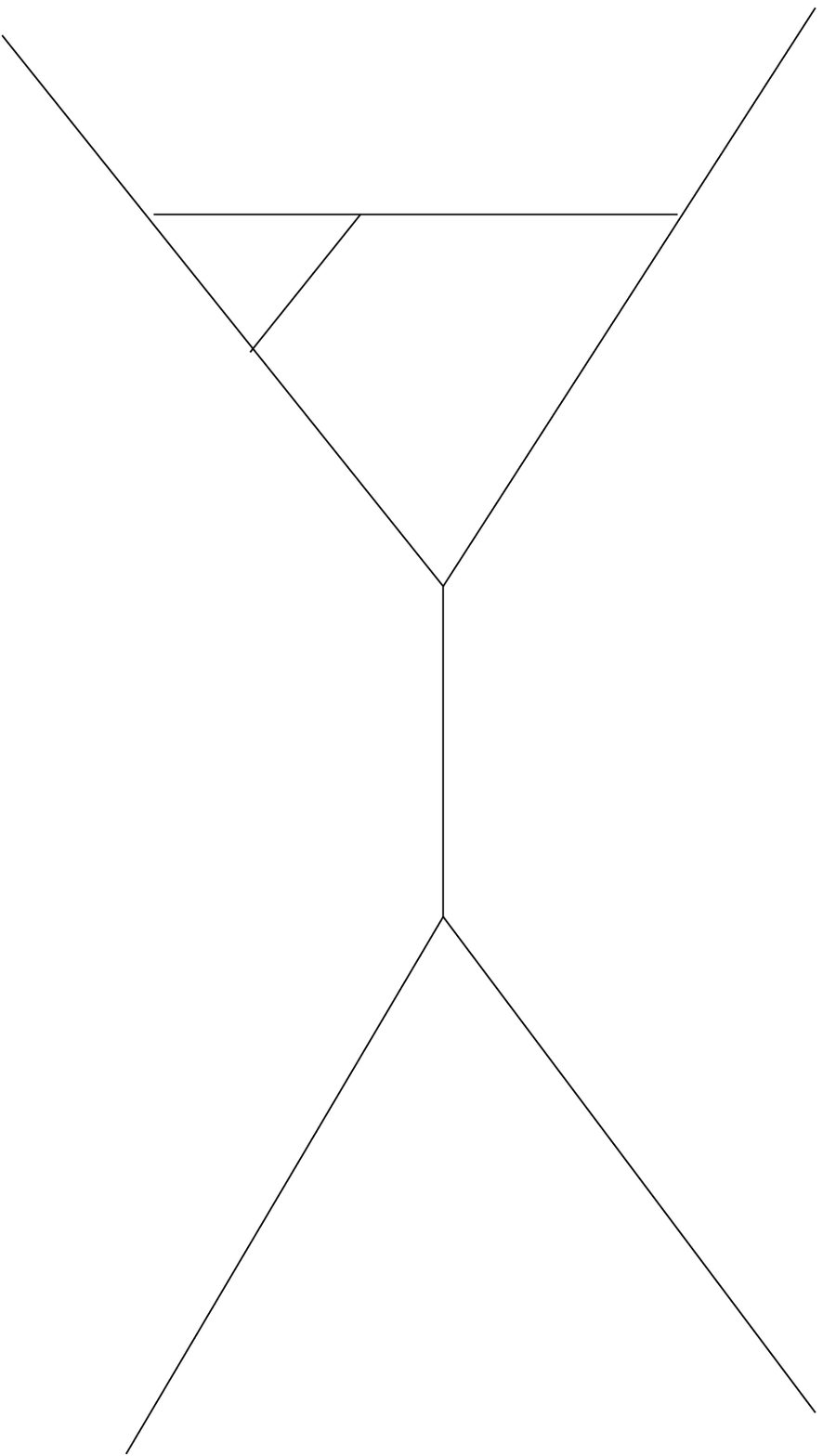, scale=0.1}}\end{array} \rp
= \ba{c}\mbox{\fig{file=c1.eps, scale=0.07}}\ea \otimes \begin{array}{c}\mbox{\epsfig{file=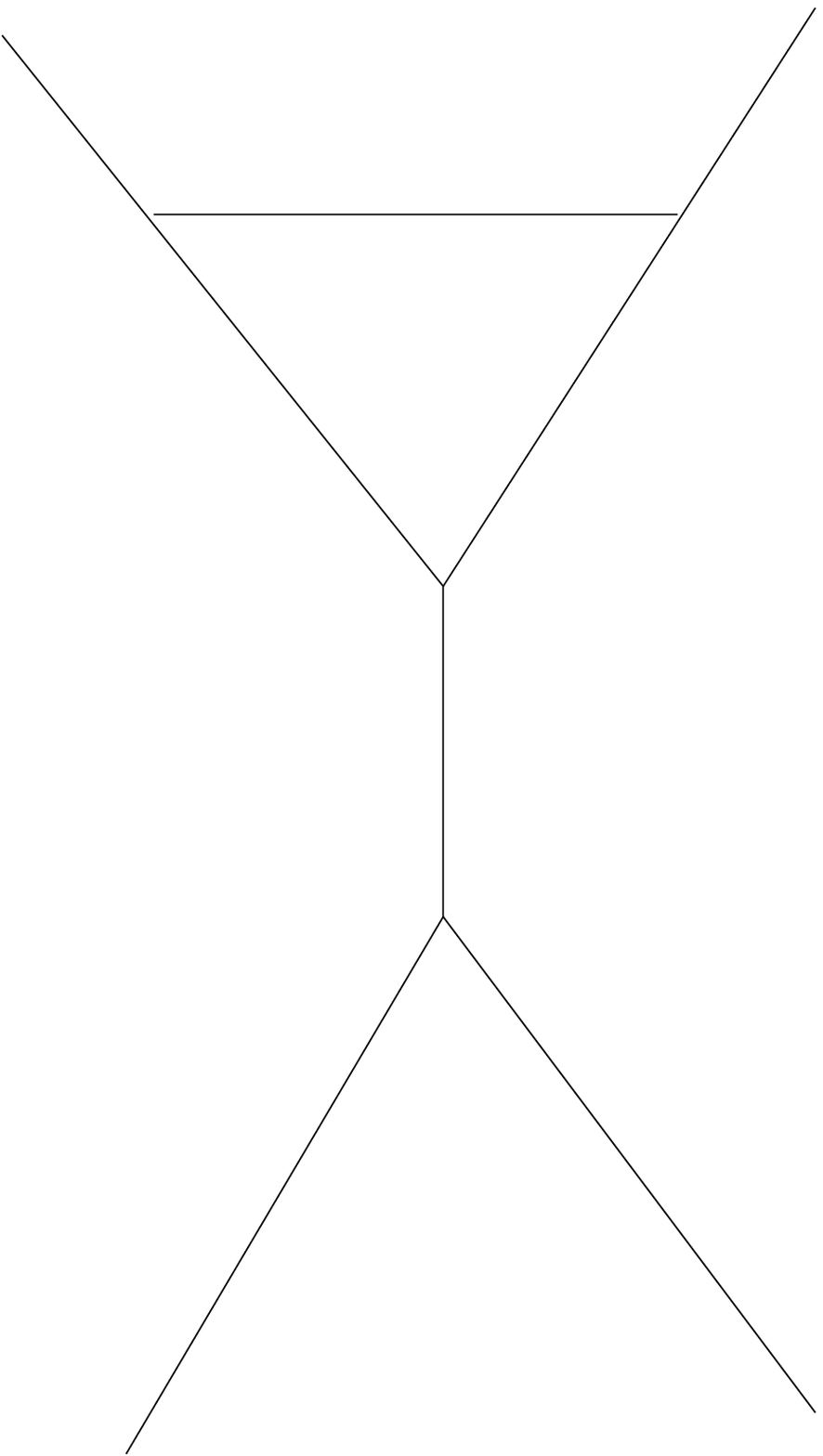, scale=0.1}}\end{array} +
 \ba{c}\mbox{\fig{file=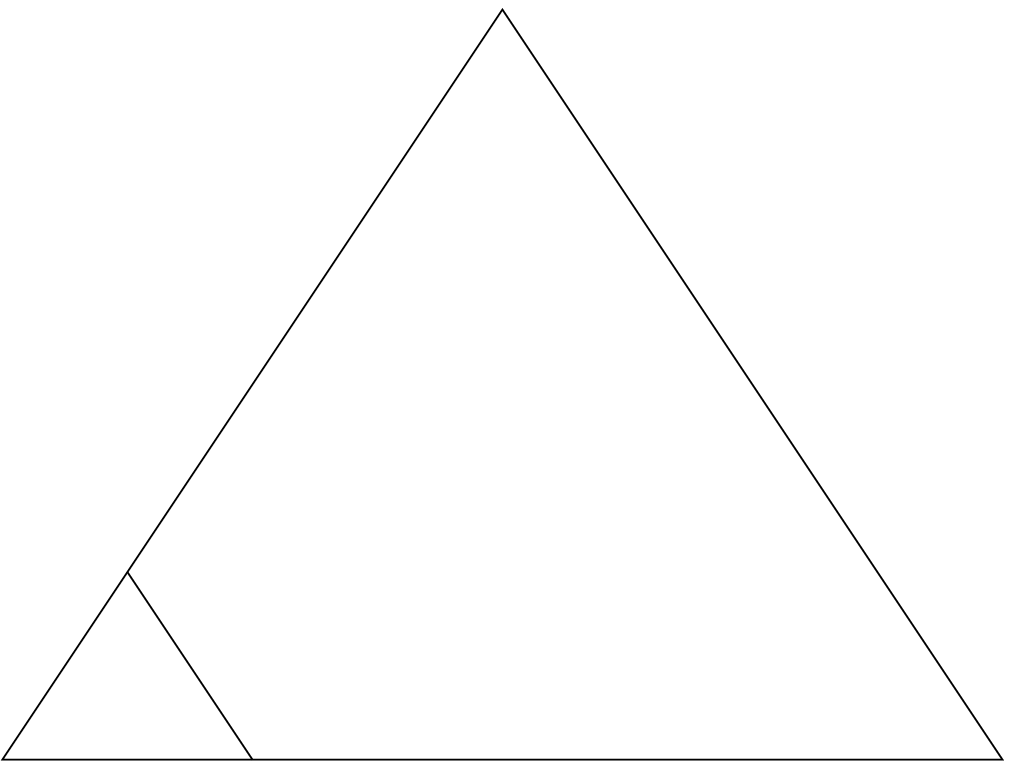, scale=0.07}}\ea \otimes \begin{array}{c}\mbox{\epsfig{file=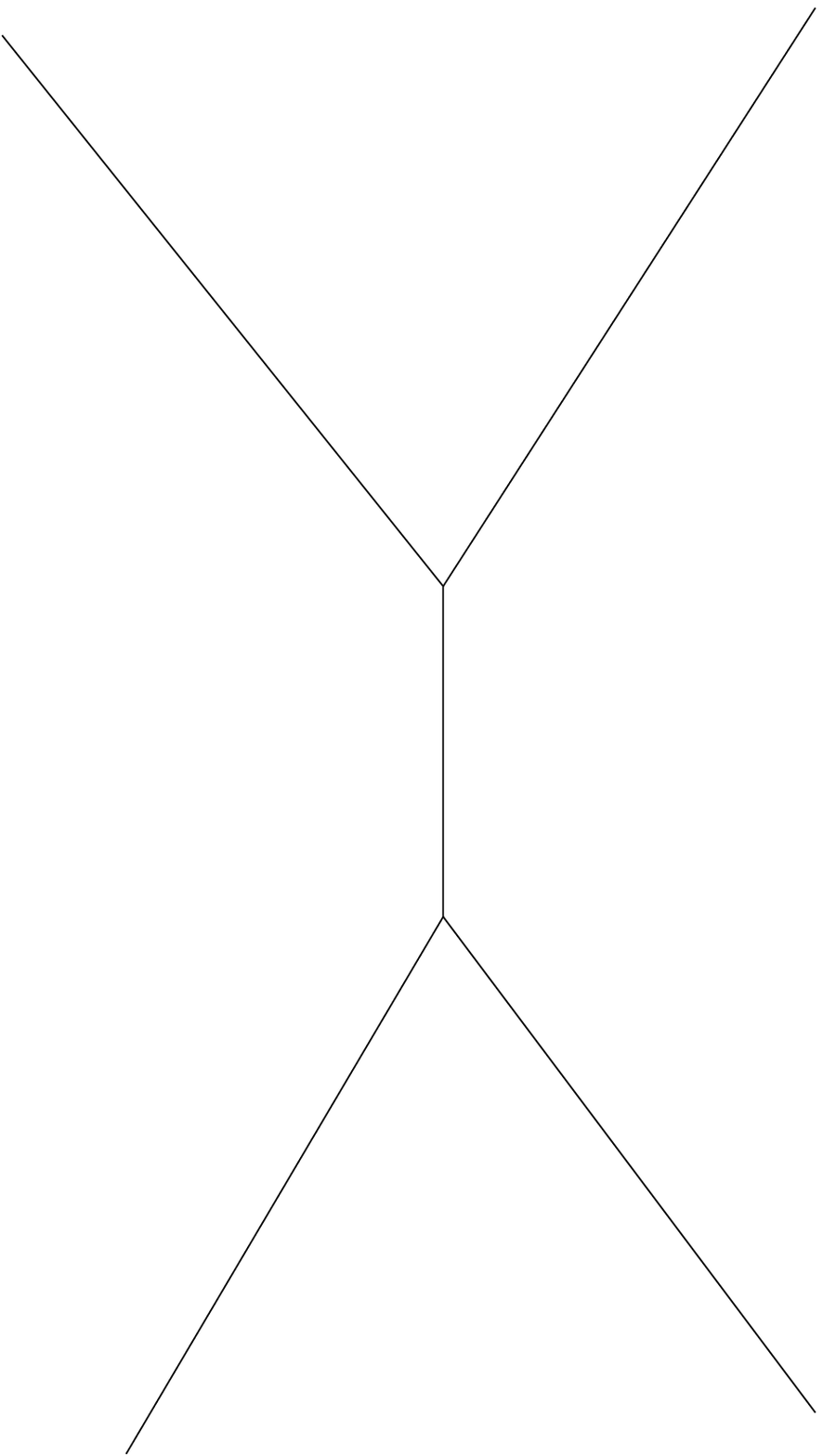, scale=0.1}}\end{array}\, .
\eeqa
Note that, unlike \cite{fm}, we do not use a white vertex (or any other graphical object) to remind where the shrinking of the sub-SF was done in the SF.
A crucial observation is that, in the algebraic construction of \cite{fm} one allows ``tree''-like elements such as
\beqa
\mbox{\epsfig{file=fm-tree.eps, scale=0.1}}\, .
\eeqa
Nevertheless, this type of SFs cannot be obtained on the LHS when acting with the coproduct, unless some supplementary notion (like some kind of ``color'' of the SFs) is defined. In order to obtain the same properties as in QFT, we need for this  to be satisfied also. This will become clear in the following.
We propose here to {\it quotient out} this sector. Note that these ``tree''-like SFs form a trivial Hopf coideal. For the sake of completeness let us also remark that they do not form a Hopf ideal.

We denote the quotiented structure by $\mathbb T$ and we refer to it as the {\it core Hopf algebra of SFs}, for the reasons explained above. 
To check that $\mathbb T$ is a Hopf algebra one has just make the correspondence with the Hopf algebra of rooted trees \cite{rt, rt-ck}. This correspondence is immediate. 
The graduation of  $\mathbb T$ is given naturally by the number of triangles of the respective SF. In the language of rooted trees, this corresponds to the weight of the tree (the number of vertices of the respective tree).

The empty SF is denoted by $1_{\mathbb T}$ and is the only element of the algebra of graduation $0$. For graduation $1$ one has the SF
$$\ba{c}\mbox{\fig{file=c1.eps, scale=0.07}}\ea\, .$$
For graduation $2$ one has the SFs:
$$  \ba{c}\mbox{\fig{file=c2-1.eps, scale=0.07}}\ea\, , \ba{c}\mbox{\fig{file=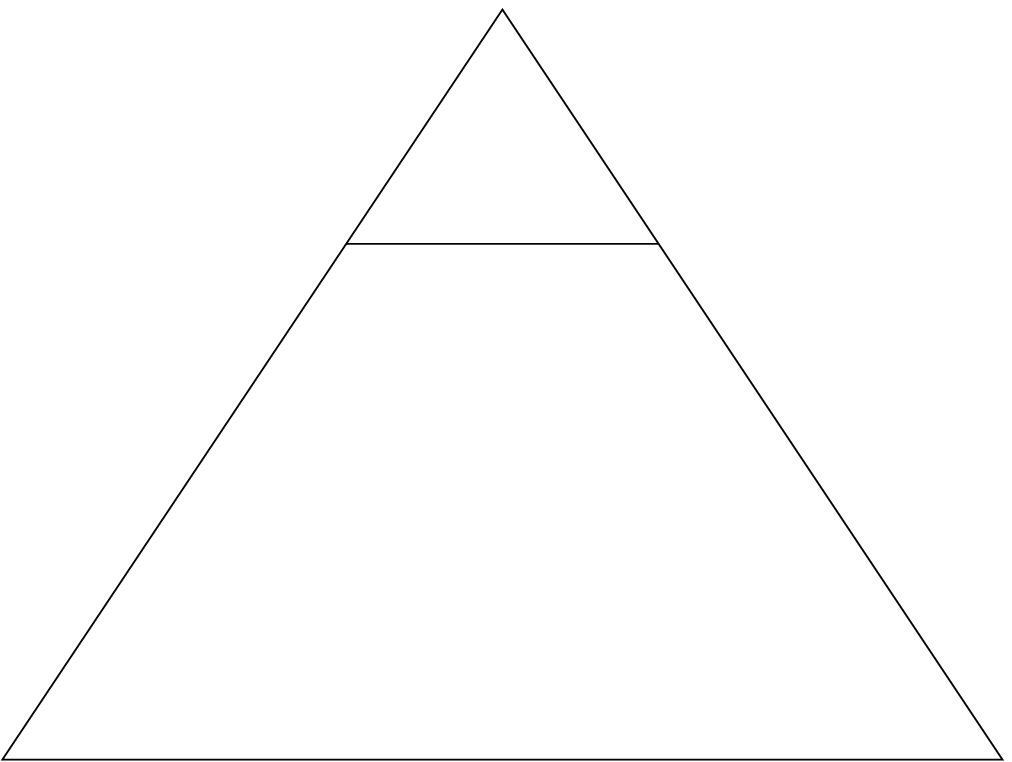, scale=0.07}}\ea\,  , \ba{c}\mbox{\fig{file=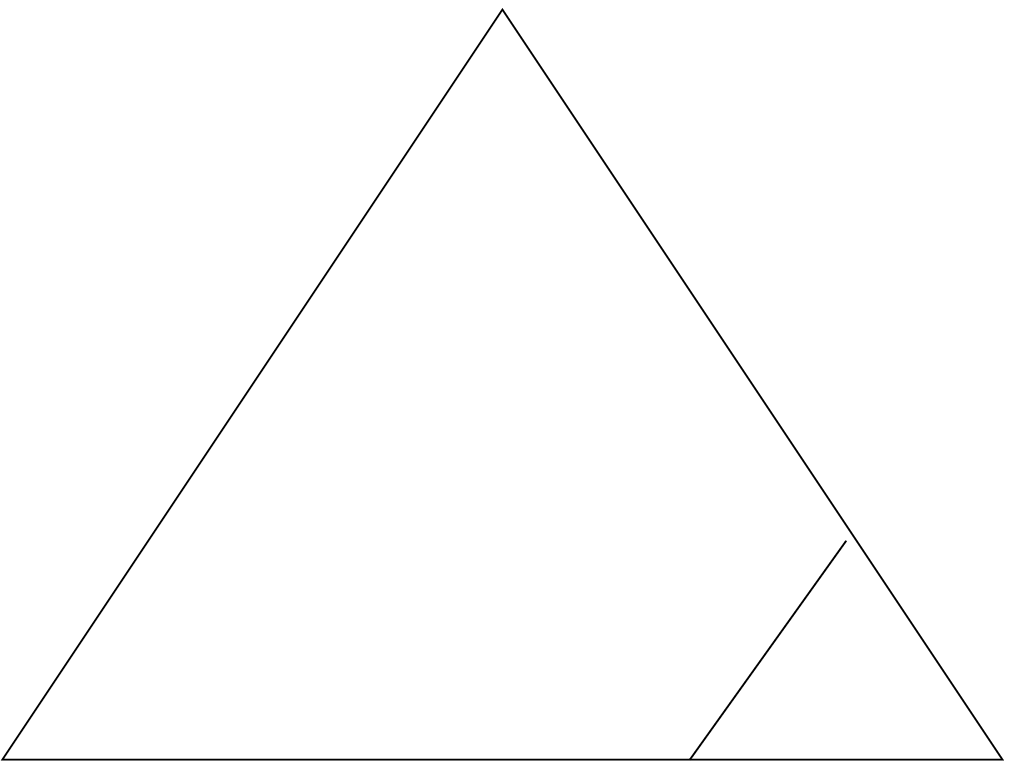, scale=0.07}}\ea\, .$$
The graduation $3$ ones are
$$  \ba{c}\mbox{\fig{file=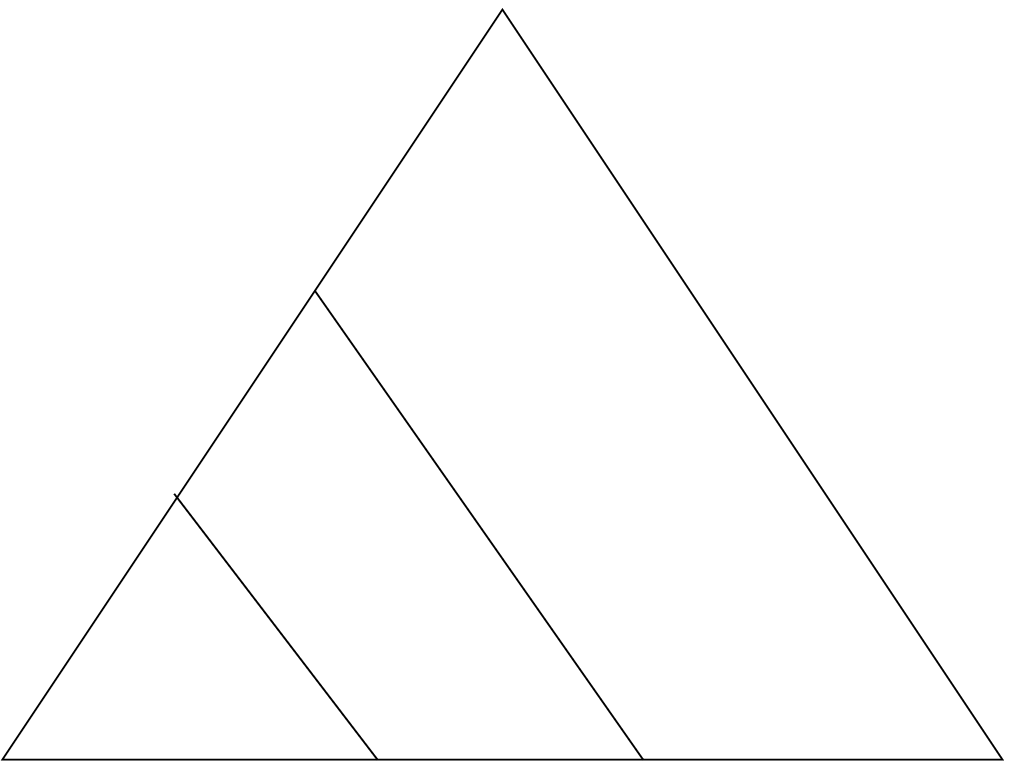, scale=0.07}}\ea\, , \ba{c}\mbox{\fig{file=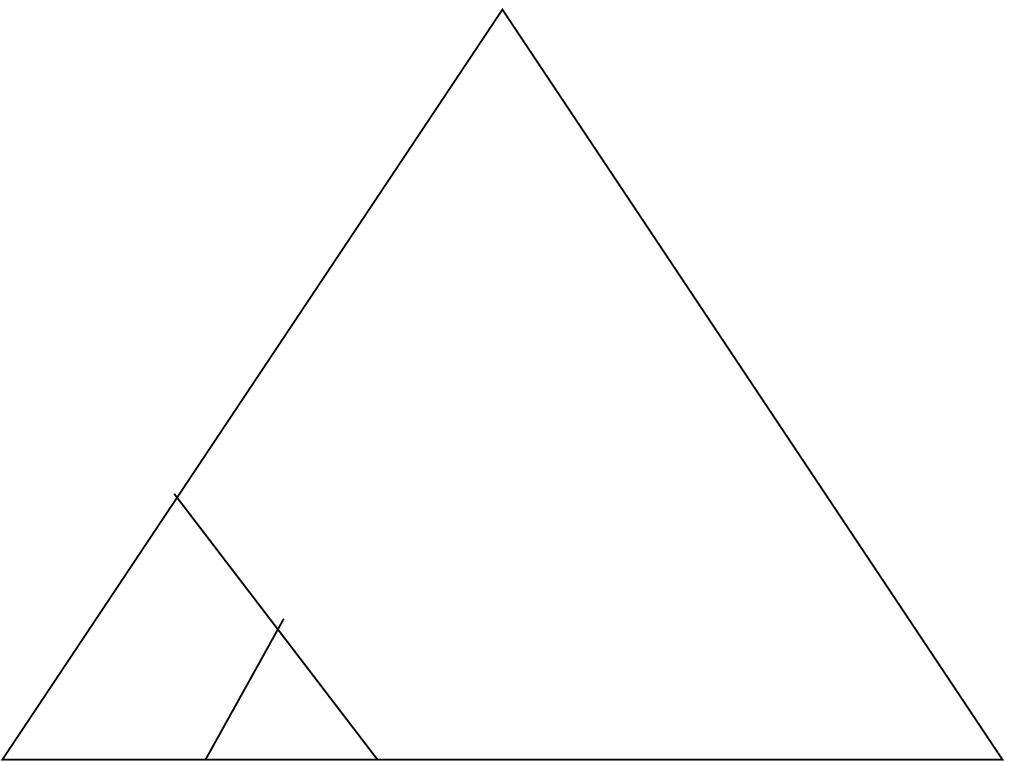, scale=0.07}}\ea\,  , \ba{c}\mbox{\fig{file=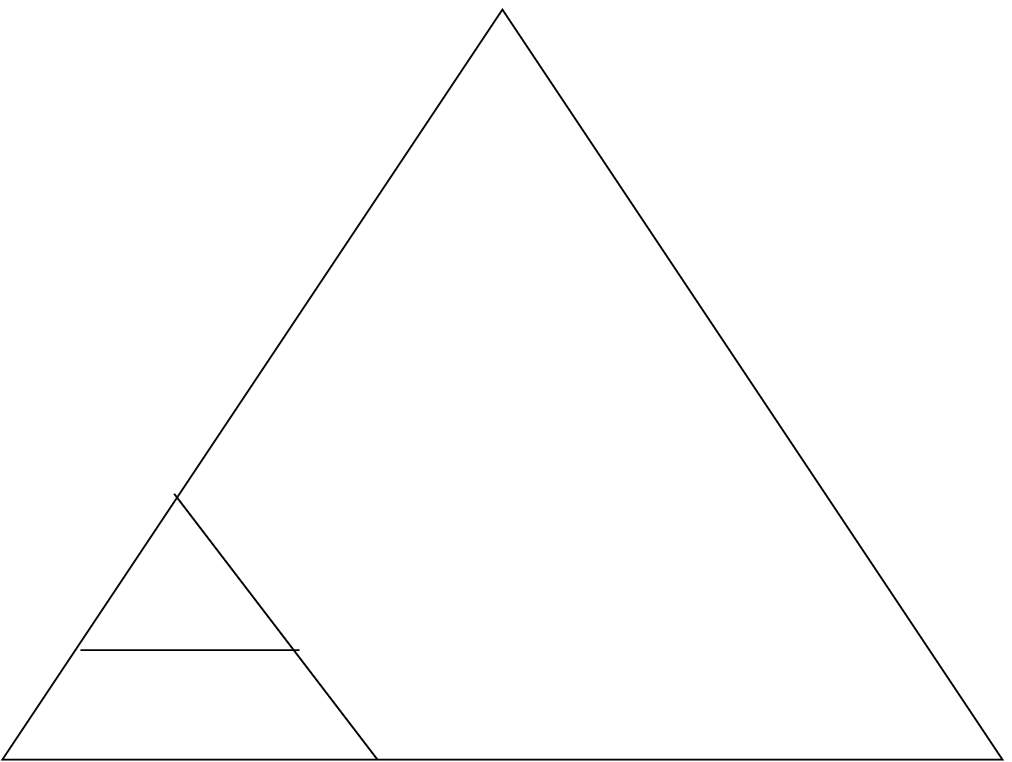, scale=0.07}}\ea\, ,
 \ba{c}\mbox{\fig{file=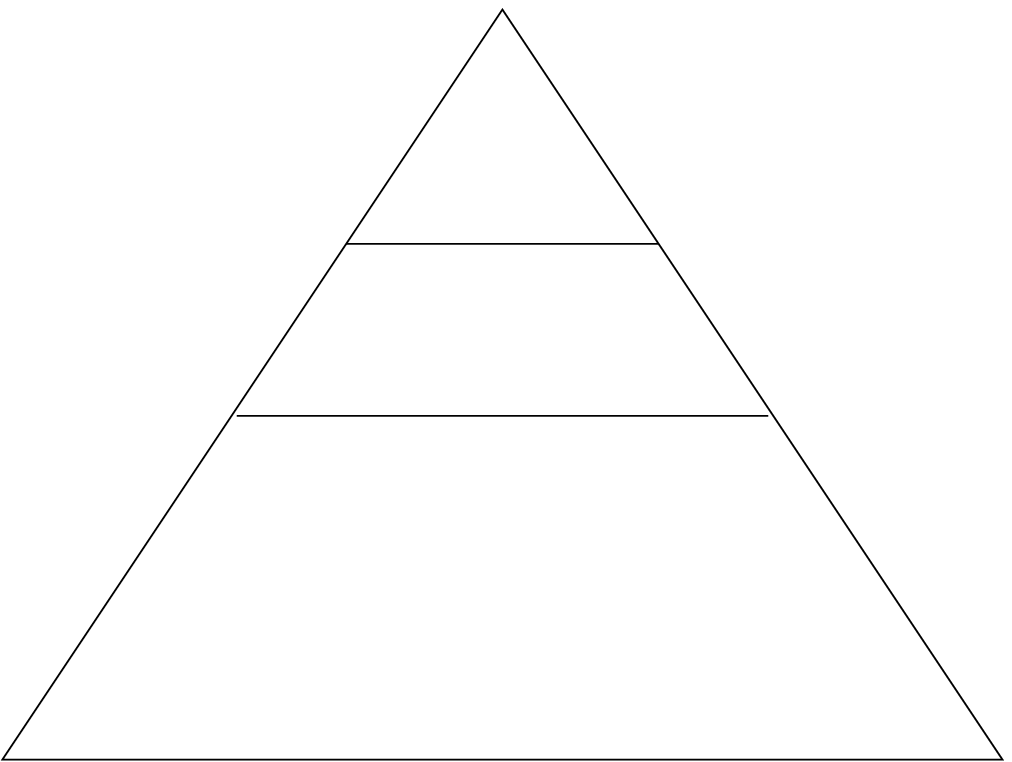, scale=0.07}}\ea\, , \ba{c}\mbox{\fig{file=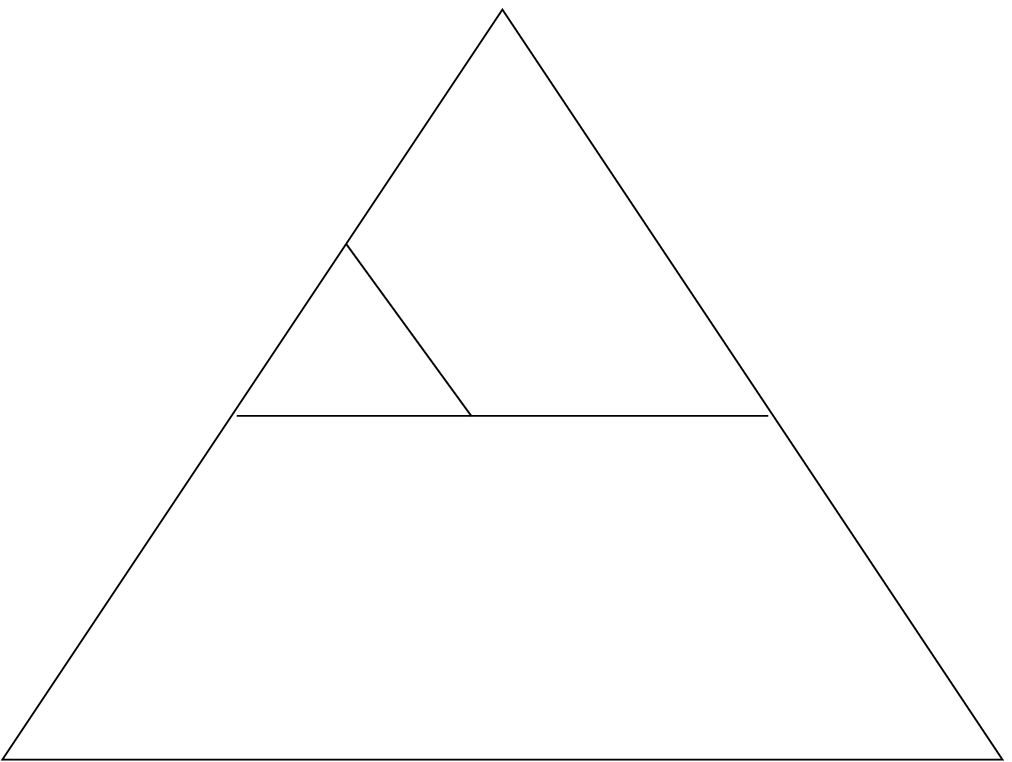, scale=0.07}}\ea\,  , \ba{c}\mbox{\fig{file=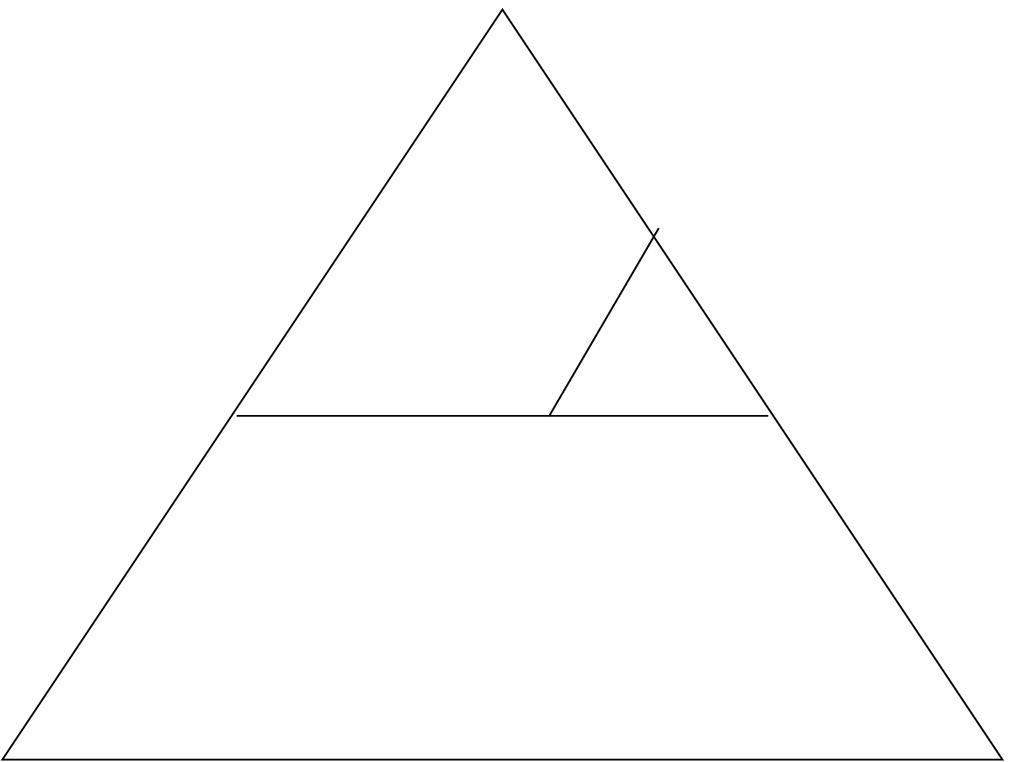, scale=0.07}}\ea\, ,
 \ba{c}\mbox{\fig{file=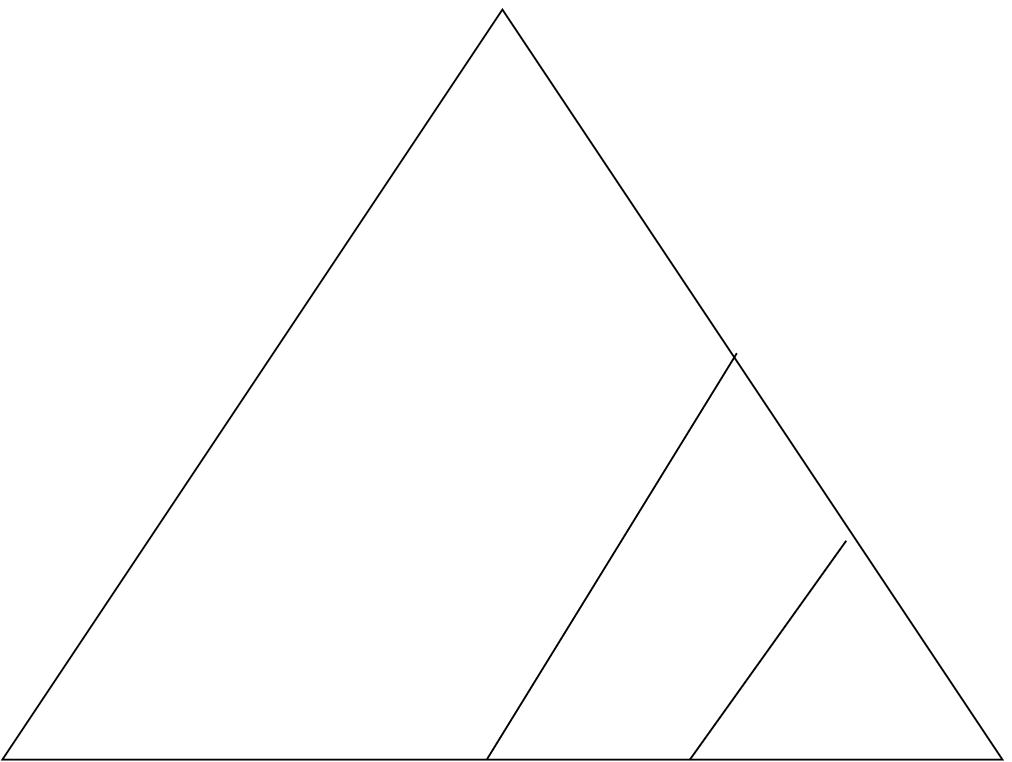, scale=0.07}}\ea\, , \ba{c}\mbox{\fig{file=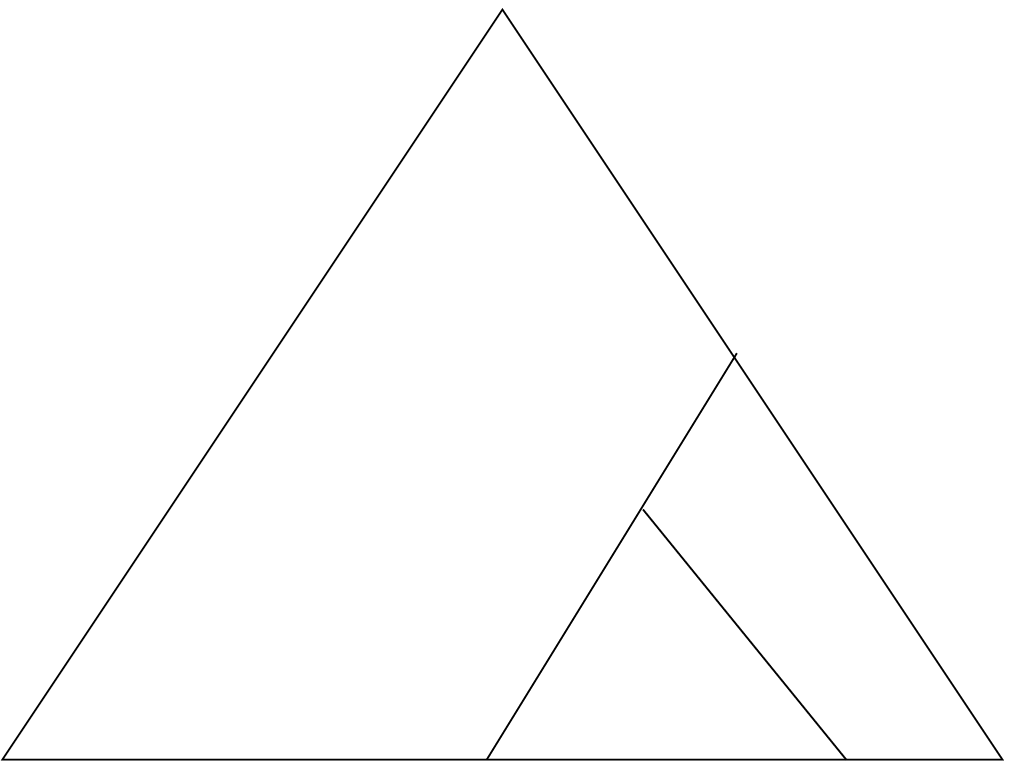, scale=0.07}}\ea\,  , \ba{c}\mbox{\fig{file=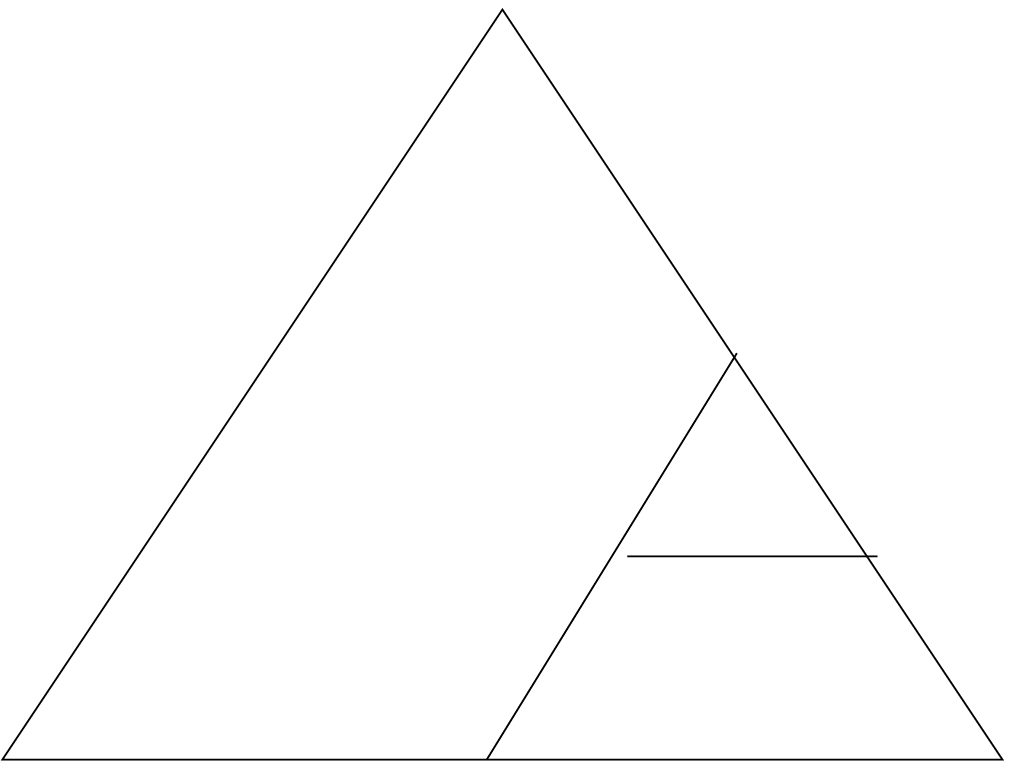, scale=0.07}}\ea\, ,
 \ba{c}\mbox{\fig{file=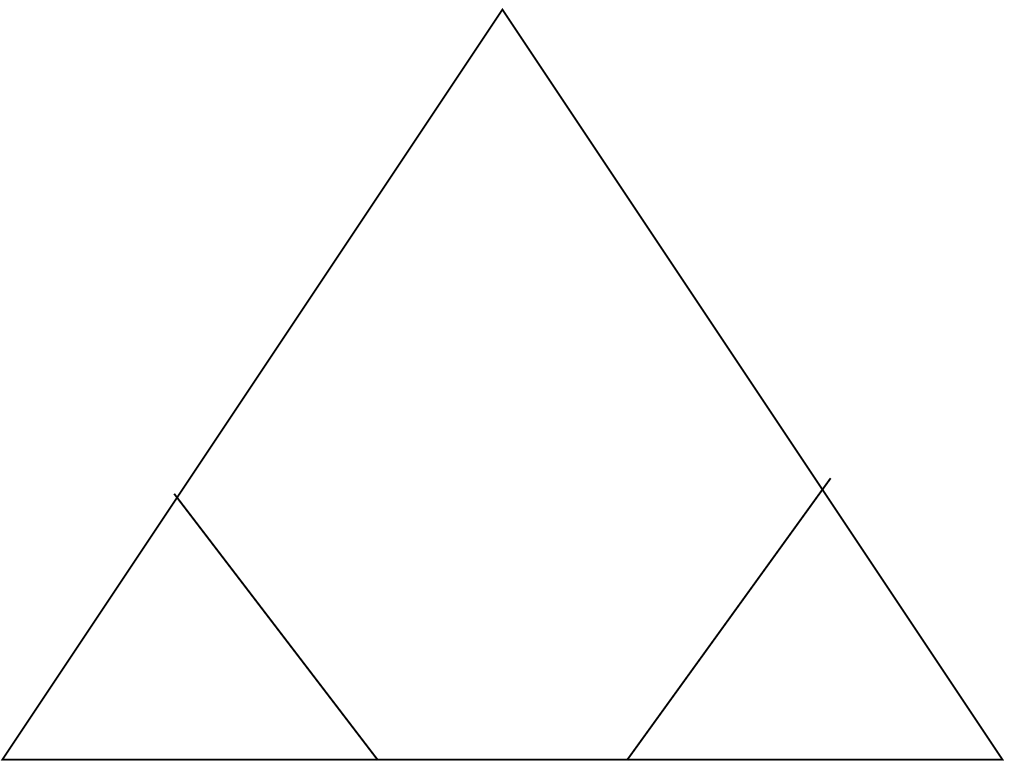, scale=0.07}}\ea\, , \ba{c}\mbox{\fig{file=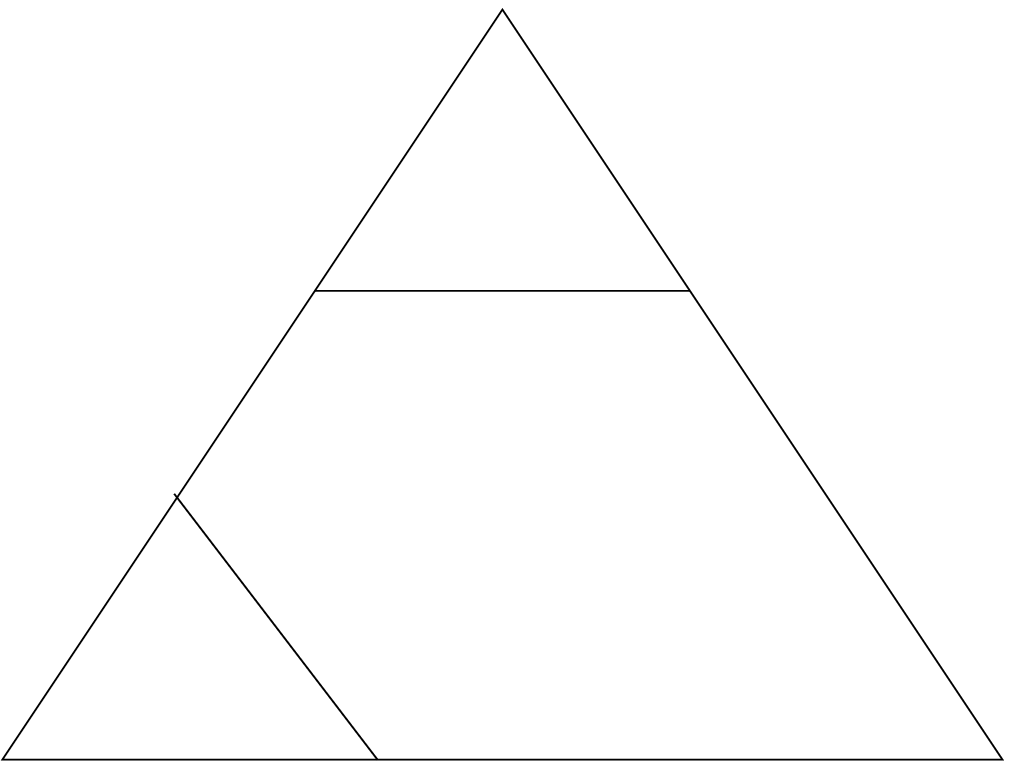, scale=0.07}}\ea\,  , \ba{c}\mbox{\fig{file=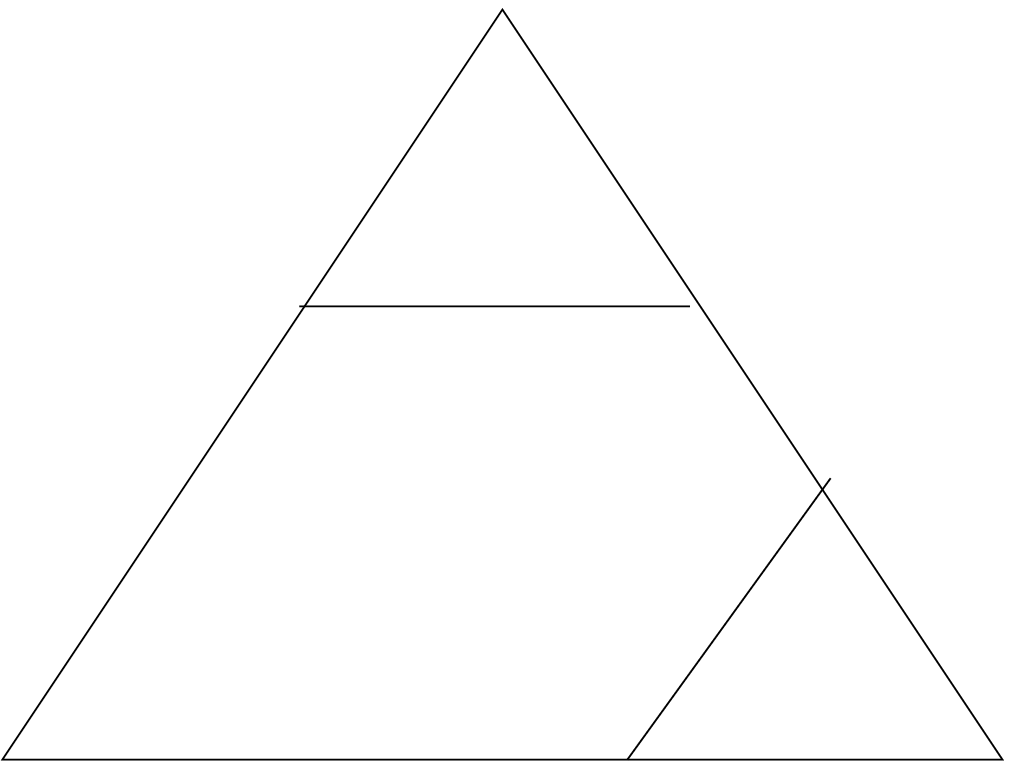, scale=0.07}}\ea$$
and so on.

We denote the coproduct by $\Delta_\T$:
\beqa
\label{coproduct}
\Delta_T \Gamma= \Gamma \otimes \unu + \unu \otimes \Gamma + \sum_{\gamma\subset\Gamma} \gamma\otimes \Gamma/\gamma,
\eeqa
where, as in \eqref{coproductM}, $\Gamma$ is some SF and $\gamma$ any of its SFs. We denote the non-trivial part of this coproduct by $\Delta'_T$.

The multiplication is, as in \cite{fm}, the disjoint union. The rest of the operations are also defined as in \cite{fm}.

Note that the rooted tree Hopf algebra has been extensively studied in recent mathematical literature (see for example \cite{foissy1, foissy2} and references within). 


\medskip

We now define a {\it grafting operator} 
$ B_+:\T\to\T$
 which increases the graduation by one unit by inserting the respective SF into a bigger triangle. Note that one has three distinct insertion places, corresponding to the three corners of the triangle. One has 
\beqa
\label{b1}
B_+ \left( \includegraphics[scale=0.07]{c1.eps} \right)=\frac 13
\left( \includegraphics[scale=0.07]{c2-1.eps}+ \includegraphics[scale=0.07]{c2-2.eps}+ \includegraphics[scale=0.07]{c2-3.eps} \right).
\eeqa 
Note that the internal structure ({\it i. e.} internal triangles) does not play when acting with the grafting operator. Furthermore, one has
\beqa
\label{b2}
B_+ (\tri \, \tri) &=& \frac 13 \lp 
\includegraphics[scale=0.11]{c3-10.eps}+\includegraphics[scale=0.11]{c3-11.eps}
+ \includegraphics[scale=0.11]{c3-12.eps}\rp \, ,\nonumber\\
B_+ (\tri \, \tri\, \tri)&=& \includegraphics[scale=0.07]{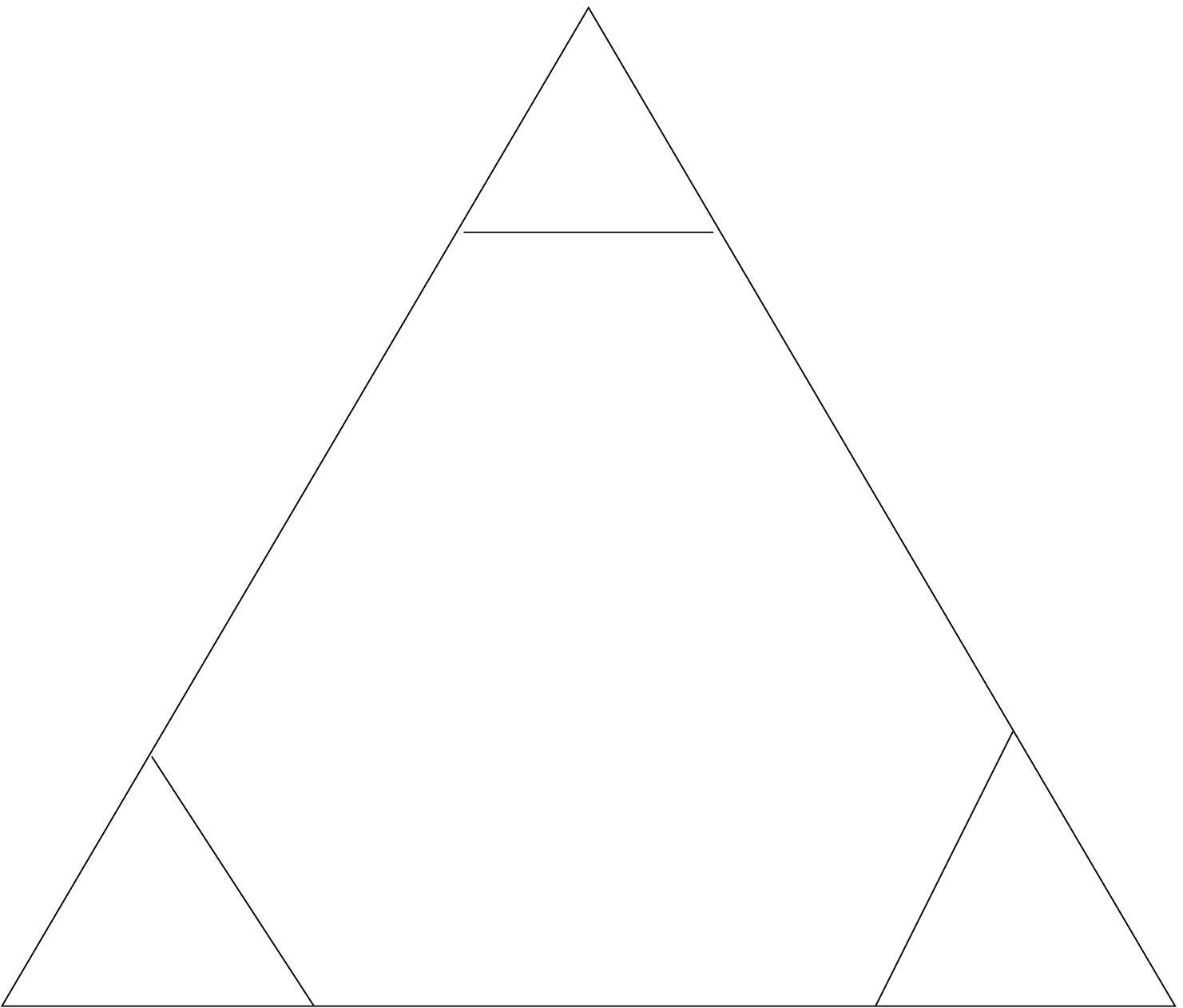}\, ,\nonumber\\
B_+ (\tri \, \tri\, \tri\, \tri\, \ldots)&=&0,
\eeqa
where, by $\ldots$ in the last line above, we mean any number ($0$ included) of $\tri$. This comes from the fact that we work with three maximal insertion places. This is related, in the rooted tree language, to the fertility of a vertex of a tree, that is the number of outgoing edges (see for example \cite{bk}).
To complete the definition, one has
\beqa
\label{trivial}
B_+ (1_{\mathbb T})=\includegraphics[scale=0.07]{c1.eps}.
\eeqa
 The naturality of these equations will become clear in the following (see equations \eqref{frumusete}).
The operator $B_+$ is, from a mathematical point of view, a Hochschild one-cocycle \cite{rt-ck}.




\medskip

Let us now write down the following equation in ${\mathbb T}[[t]]$
\beqa
\label{DSE}
X=1_{\mathbb T}+t B_+ (X^3),
\eeqa
$t$ being a parameter which counts the number of triangles (this is the equivalent of the parameter counting the number of loops in the Feynman graph Connes-Kreimer algebra of renormalization).
Using the ansatz
\beqa
\label{ansatz}
X=\sum_{n=0}^\infty t^n c_n,
\eeqa
one can determine $X$ by induction. In the QFT language, this equation is nothing but a cubic combinatorial Dyson-Schwinger equation. Nevertheless, there are some differences with the combinatorial Dyson-Schwinger equations generally used in the rooted tree framework; this implies important differences in the  results, which are to be obtained in the rest of this section (see the discussion of the end of the following section).
Here we deal with such a cubic equation, because the maximal number of insertion places is three, as already stated above.

Equations \eqref{DSE} and \eqref{ansatz}, allow one 
to obtain the following results (at the first four orders in the development in the constant $t$):
\beqa
\label{frumusete}
c_0&=&\unu,\nonumber\\
c_1&=&B_+(c_0^3)=B_+(c_0)=B_+(\unu)=\includegraphics[scale=0.07]{c1.eps}\, ,\nonumber\\
c_2&=&3B_+(c_0^2c_1)=3B_+(c_1)=3B_+\left(\includegraphics[scale=0.07]{c1.eps}\right)=\includegraphics[scale=0.07]{c2-1.eps}+\includegraphics[scale=0.07]{c2-2.eps}+\includegraphics[scale=0.07]{c2-3.eps}\, ,\nonumber\\
c_3&=&3B_+(c_0c_1^2+c_0^2c_2)=3B_+(c_1^2+c_2)
=3B_+\left(\includegraphics[scale=0.07]{c1.eps}\, \includegraphics[scale=0.07]{c1.eps}+\includegraphics[scale=0.07]{c2-1.eps}+\includegraphics[scale=0.07]{c2-2.eps}+\includegraphics[scale=0.07]{c2-3.eps}\right)\nonumber\\
&=&\includegraphics[scale=0.11]{c3-1.eps}+ \includegraphics[scale=0.11]{c3-2.eps}+ \includegraphics[scale=0.11]{c3-3.eps}+\includegraphics[scale=0.11]{c3-4.eps}+\includegraphics[scale=0.11]{c3-5.eps}+ \includegraphics[scale=0.11]{c3-6.eps}\nonumber\\
&+&\includegraphics[scale=0.11]{c3-7.eps}+\includegraphics[scale=0.11]{c3-8.eps}+ \includegraphics[scale=0.11]{c3-9.eps}+\includegraphics[scale=0.11]{c3-10.eps}+\includegraphics[scale=0.11]{c3-11.eps}
+ \includegraphics[scale=0.11]{c3-12.eps}\, .
\eeqa
where we have used \eqref{b1} and \eqref{b2}. 

In all generality, using the Newton binomial formula, one proves
\beqa
\label{newton}
c_{n+1}=\sum_{k_1+ k_2+k_3=n} B_+(c_{k_1}c_{k_2}c_{k_3}).
\eeqa

\medskip

All this allows one to state the following results:
\beqa
\label{toate1}
\Delta_T(B_+)=
B_+\otimes 1_{\mathbb T}
+({\rm id}_{{\mathbb T}}\otimes B_+)\Delta_T
\eeqa
and
\beqa 
\label{toate2}
\Delta_T(c^{}_{n})=\sum_{k=0}^n P^n_k\otimes
c^{}_{k},
\eeqa 
where $P_k^n$ is a polynomial in the
variables $c_\ell$, $\ell\le n$ of total degree $n-k$.

\medskip

The proof of these identities is straightforward, being a consequence of the fact that, as stated above, one has a direct correspondence between the Hopf algebra $\mathbb T$ and the Hopf algebra of rooted trees. Thus, the identity \eqref{toate1} is in mathematical language the translation of the fact that the operator $B_+$ is a Hochschild one-cocycle \cite{bk}. The identity \eqref{toate2} is a consequence of \eqref{toate1} and can be proved by induction. In \cite{bk} such a proof was given for the combinatorial Dyson-Schwinger equation \eqref{DSE-k}. Let us 
give a proof for
our case.

For $n=0$, identity \eqref{toate2} is trivially satisfied. We now start our induction. Using \eqref{newton}, one writes:
\beqa
\Delta_\T c_n = \Delta_\T \sum_{k_1+k_2+k_3=n-1} B_+ \lp c_{k_1}c_{k_2}c_{k_3}\rp.
\eeqa
We now make use of \eqref{toate1} to obtain:
\beqa
\label{inter}
\Delta_\T c_n= \sum_{k_1+k_2+k_3=n-1}  B_+ \lp c_{k_1}c_{k_2}c_{k_3}\rp \otimes \unu + 
({\rm id}_{{\mathbb T}}\otimes B_+)\Delta_T \lp \sum_{k_1+k_2+k_3=n-1}c_{k_1}c_{k_2}c_{k_3}\rp.
\eeqa
Using again \eqref{newton}, equation \eqref{inter} becomes:
\beqa
\label{inter2}
\Delta_\T c_n= c_n \otimes \unu + 
({\rm id}_{{\mathbb T}}\otimes B_+)\Delta_T \lp \sum_{k_1+k_2+k_3=n-1}c_{k_1}c_{k_2}c_{k_3}\rp.
\eeqa
Making now use of the induction hypothesis, one has:
\beqa
\label{inter3}
\Delta_\T c_n= c_n \otimes \unu + 
({\rm id}_{{\mathbb T}}\otimes B_+)\sum_{k_1+k_2+k_3=n-1}\sum_{\ell_1,\ell_2,\ell_3}P^{k_1}_{\ell_1}P^{k_2}_{\ell_2}P^{k_3}_{\ell_3}\otimes c_{\ell_1}c_{\ell_2}c_{\ell_3},
\eeqa
which further writes
\beqa
\label{inter4}
\Delta_\T c_n= c_n \otimes \unu + 
\sum_{k_1+k_2+k_3=n-1}\sum_{\ell_1,\ell_2,\ell_3}P^{k_1}_{\ell_1}P^{k_2}_{\ell_2}P^{k_3}_{\ell_3}\otimes B_+\lp c_{\ell_1}c_{\ell_2}c_{\ell_3}\rp.
\eeqa
By rearranging the indices of the last term above
the left hand tensor factor gives $P_k^n$ and 
the right hand tensor factor, once again using \eqref{newton}, gives $c_q$ ($q=1,\ldots , n$):
\beqa
\label{inter5}
\Delta_\T c_n= c_n \otimes \unu + 
\sum_{q=1}^n P^n_q\otimes c_q.
\eeqa
By a direct inspection, one can see that $P_q^n$ is nothing more then an homogeneous polynomial in the variables $c_\ell$ ($\ell\le n$) of total degree $n-q$. Furthermore, let us recall that $P^n_0=c_n$.

\medskip

The identity \eqref{toate2} thus shows that the elements $c_n$ form Hopf subalgebras in $\T$. In QFT, this type of result is of fundamental importance for finding some exact solutions of the Dyson-Schwinger equations \cite{exact}.


\medskip

To end this section, let us illustrate identities \eqref{toate1} and \eqref{toate2} on some non-trivial particular cases of small graduation SFs.
The LHS of \eqref{toate1} applied for $c_1$ gives
\beqa
\label{l}
\Delta B_+ \lp \includegraphics[scale=0.07]{c1.eps}\rp&=&\Delta\lp \frac 13 \lp \includegraphics[scale=0.07]{c2-1.eps}+\includegraphics[scale=0.07]{c2-2.eps}+\includegraphics[scale=0.07]{c2-3.eps}\rp\rp\\
&= &\frac 13 \lp \includegraphics[scale=0.07]{c2-1.eps}+\includegraphics[scale=0.07]{c2-2.eps}+\includegraphics[scale=0.07]{c2-3.eps}\rp
\otimes1_{{\mathbb T}}+
1_{{\mathbb T}}\otimes \frac 13 \lp \includegraphics[scale=0.07]{c2-1.eps}+\includegraphics[scale=0.07]{c2-2.eps}+\includegraphics[scale=0.07]{c2-3.eps}\rp+
\includegraphics[scale=0.07]{c1.eps}\otimes \includegraphics[scale=0.07]{c1.eps}\ .\nonumber
\eeqa
On the RHS, one has
\beqa
&&B_+ (\includegraphics[scale=0.07]{c1.eps})\otimes 1_{{\mathbb T}}
+ ({\rm id}_{{\mathbb T}}\otimes B_+)\Delta(\includegraphics[scale=0.07]{c1.eps}))\\
&&=\frac 13 \lp \includegraphics[scale=0.07]{c2-1.eps}+\includegraphics[scale=0.07]{c2-2.eps}+\includegraphics[scale=0.07]{c2-3.eps}\rp
\otimes1_{{\mathbb T}}+
\includegraphics[scale=0.07]{c1.eps}\otimes \includegraphics[scale=0.07]{c1.eps}+
1_{{\mathbb T}}\otimes \frac 13 \lp \includegraphics[scale=0.07]{c2-1.eps}+\includegraphics[scale=0.07]{c2-2.eps}+\includegraphics[scale=0.07]{c2-3.eps}\rp
\ ,\nonumber
\eeqa
which is identical to \eqref{l}, as expected. The polynomials given by formula \eqref{toate2} are trivial.


Let us now go further and verify the identity \eqref{toate1} for $c_2$ given by \eqref{frumusete}. The non-trivial part of the LHS writes
\beqa
\label{l2}
\frac 13 \Delta'_\T &&\lp 
\includegraphics[scale=0.11]{c3-1.eps}+ \includegraphics[scale=0.11]{c3-2.eps}+ \includegraphics[scale=0.11]{c3-3.eps}\right.\nonumber\\
&&+\includegraphics[scale=0.11]{c3-4.eps}+\includegraphics[scale=0.11]{c3-5.eps}+ \includegraphics[scale=0.11]{c3-6.eps}\nonumber\\
&&\left.+\includegraphics[scale=0.11]{c3-7.eps}+\includegraphics[scale=0.11]{c3-8.eps}+ \includegraphics[scale=0.11]{c3-9.eps}\right).
\eeqa
This further gives the following six terms
\beqa
\label{l22}
\includegraphics[scale=0.07]{c1.eps}\otimes \lp \includegraphics[scale=0.09]{c2-1.eps}+\includegraphics[scale=0.09]{c2-2.eps}+\includegraphics[scale=0.09]{c2-3.eps}\rp,\nonumber\\ 
\lp \includegraphics[scale=0.09]{c2-1.eps}+\includegraphics[scale=0.09]{c2-2.eps}+\includegraphics[scale=0.09]{c2-3.eps}\rp \otimes \includegraphics[scale=0.07]{c1.eps}.
\eeqa
On the RHS, the non-trivial terms are obtained from
\beqa
\label{r2}
({\rm id}_{{\mathbb T}}\otimes B_+)\Delta'_\T  \lp \includegraphics[scale=0.09]{c2-1.eps}+\includegraphics[scale=0.09]{c2-2.eps}+\includegraphics[scale=0.09]{c2-3.eps}\rp.
\eeqa
The non-trivial terms of \eqref{r2} are given by
\beqa
({\rm id}_{{\mathbb T}}\otimes B_+)\lp  \lp \includegraphics[scale=0.09]{c2-1.eps}+\includegraphics[scale=0.09]{c2-2.eps}+\includegraphics[scale=0.09]{c2-3.eps}\rp \otimes \unu
+3 \includegraphics[scale=0.07]{c1.eps}\otimes\includegraphics[scale=0.07]{c1.eps}\rp
\eeqa
This gives
\beqa
\includegraphics[scale=0.07]{c1.eps}\otimes \lp \includegraphics[scale=0.09]{c2-1.eps}+\includegraphics[scale=0.09]{c2-2.eps}+\includegraphics[scale=0.09]{c2-3.eps}\rp\nonumber\\
\lp \includegraphics[scale=0.09]{c2-1.eps}+\includegraphics[scale=0.09]{c2-2.eps}+\includegraphics[scale=0.09]{c2-3.eps}\rp \otimes \includegraphics[scale=0.07]{c1.eps}.
\eeqa
which are, as expected, the same six terms as in \eqref{l22}.

Let us now explicitly verify identity \eqref{toate2} at the graduation $2$ level. Considering only the non-trivial part of the coproduct, one has
\beqa
\Delta'_{{\mathbb T}} c_2 = 3  \includegraphics[scale=0.07]{c1.eps}\otimes \includegraphics[scale=0.07]{c1.eps} = 3 \includegraphics[scale=0.07]{c1.eps}\otimes c_1,
\eeqa
and thus
\beqa
P_1^2=3c_1.
\eeqa



Finally, let us now explicitly verify identity \eqref{toate2} at the graduation $3$ level. Considering the non-trivial part of the coproduct, one has
\beqa
\Delta'_\T&=& 5\,  \tri \otimes  \lp \includegraphics[scale=0.07]{c2-1.eps} + \includegraphics[scale=0.07]{c2-2.eps} + \includegraphics[scale=0.07]{c2-3.eps}\rp
+ \lp 3 \, \tri \, \tri + 3 \lp \includegraphics[scale=0.07]{c2-1.eps} + \includegraphics[scale=0.07]{c2-2.eps} + \includegraphics[scale=0.07]{c2-3.eps}\rp \rp \otimes \tri\nonumber\\
&=& 5\, \tri \otimes c_2 + \lp 3 \, \tri \, \tri + 3 \lp \includegraphics[scale=0.07]{c2-1.eps} + \includegraphics[scale=0.07]{c2-2.eps} + \includegraphics[scale=0.07]{c2-3.eps}\rp \rp \otimes c_1,
\eeqa
which leads to
\beqa
P^3_1&=&3c_1^2+3c_2,\nonumber\\
P^3_2&=&5c_1.
\eeqa
We list these results in Table \ref{tabel1}.

\begin{table}[ht]
\caption{First values of the polynomial $P^n_k$ for the $2D$ case} 
\centering 
\begin{tabular}{c c c c c} 
\hline\hline 
$P^n_k$ & n=0 & n=1 &  n=2 & n=3 \\ [0.5ex] 
\hline 
k=0 & $1_\T$ & $c_1$ & $c_2$ & $c_3$ \\ 
k=1 &  & $1_\T$ & $3c_1$ & $3c_1^2+3c_2$ \\
k=2 &  &  &  $1_\T$ & $5c_1$\\
k=3 &  &  & & $1_\T$\\
[1ex] 
\hline 
\end{tabular}
\label{tabel1} 
\end{table}

\medskip

Let us end this section be stating that, as in \cite{fm}, one can define an analogous algebraic structure on the vector space of parenthesized SFs over $\CC$.

\section{Generalization to $3D$ and $4D$}
\renewcommand{\theequation}{\thesection.\arabic{equation}}    
\setcounter{equation}{0}
\label{generalizare}

In this section, we generalize the previous results to the case of $3D$ and resp. $4D$ SFs.This generalization is rather natural, since the number of maximal insertion places goes from three (in the $2D$ case) to four (in the $3D$ case) and resp. five (in the $4D$ case).

\subsection{The $3D$ case}

\begin{figure}
\centerline{\epsfig{figure=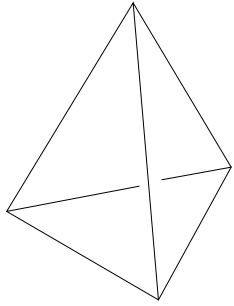,width=3cm} }
\caption{A tetrahedron is the building block of the core Hopf algebra of  $3D$ SFs. It plays the same role as the triangle in the $2D$ construction, being the graduation one generator in the algebra.}
\label{tetra}
\end{figure}

\begin{figure}
\centerline{\epsfig{figure=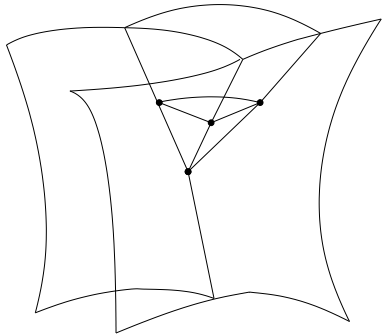,width=5cm} }
\caption{An example of a  $3D$ SF.}
\label{3dsf}
\end{figure}

In the $3D$ case, the building block which replaces the triangle is the tetrahedron of Fig. \ref{tetra}. 
These tetrahedrons are related to SFs, as shown in Fig. \ref{3dsf}.

One naturally generalizes \eqref{b1}, \eqref{b2} and \eqref{trivial} to define the grafting operator $B_+$. The equation corresponding to \eqref{DSE} is now
\beqa
\label{3DSE}
X=\unu+tB_+ (X^4).
\eeqa
Proceeding as in the previous section, one writes down (at the first four orders in the development in the constant $t$):
\beqa
\label{frumusete3}
c_0&=&\unu,\nonumber\\
c_1&=&B_+(c_0^4)=B_+(\unu)=\ba{c}\mbox{\fig{file=tetraedru.eps, scale=0.07}}\ea
\, ,\nonumber\\
c_2&=&4B_+(c_0^3c_1)=4B_+(c_1)=4B_+\left(\ba{c}\mbox{\fig{file=tetraedru.eps, scale=0.07}}\ea\right)=\ba{c}\mbox{\fig{file=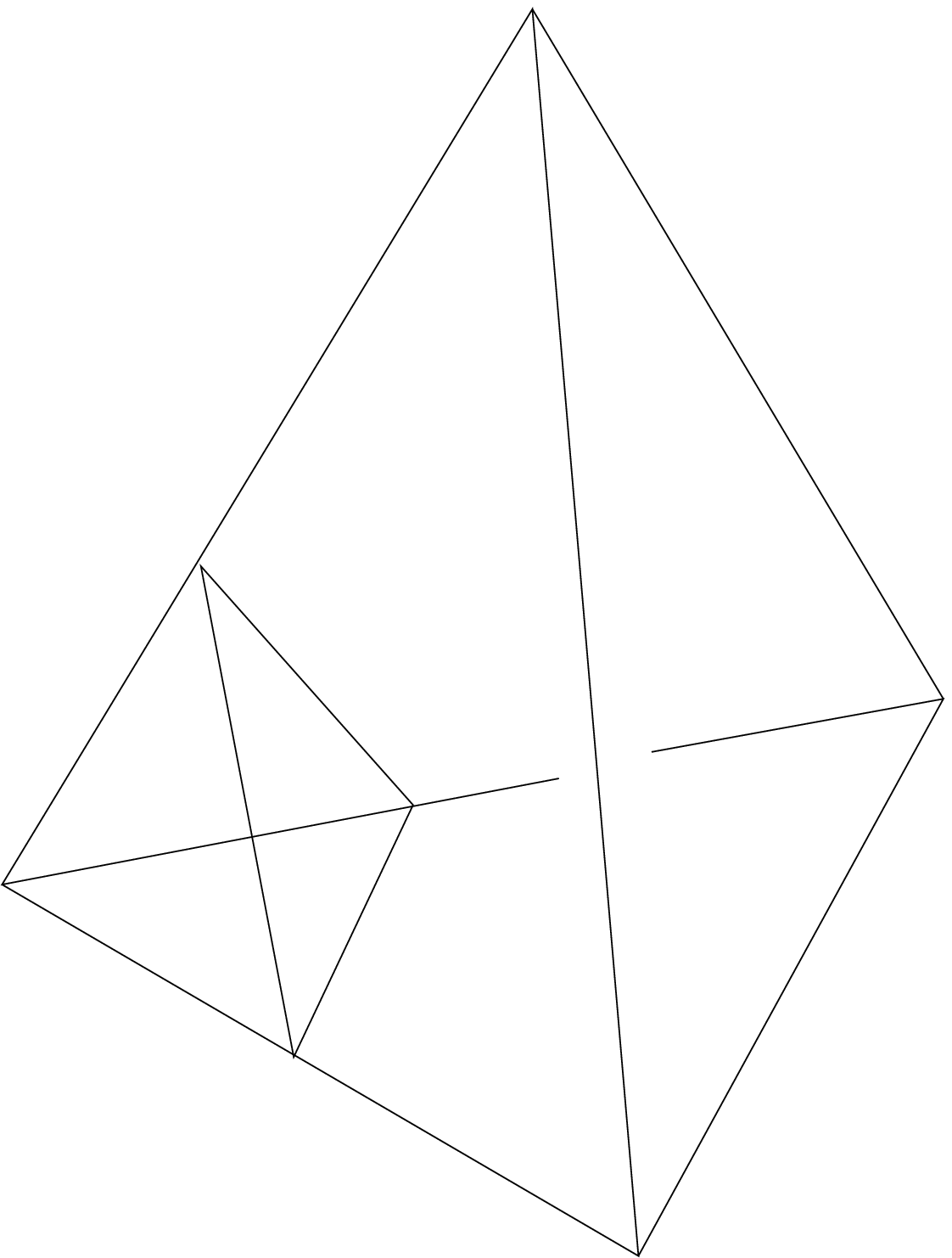, scale=0.1}}\ea+\ba{c}\mbox{\fig{file=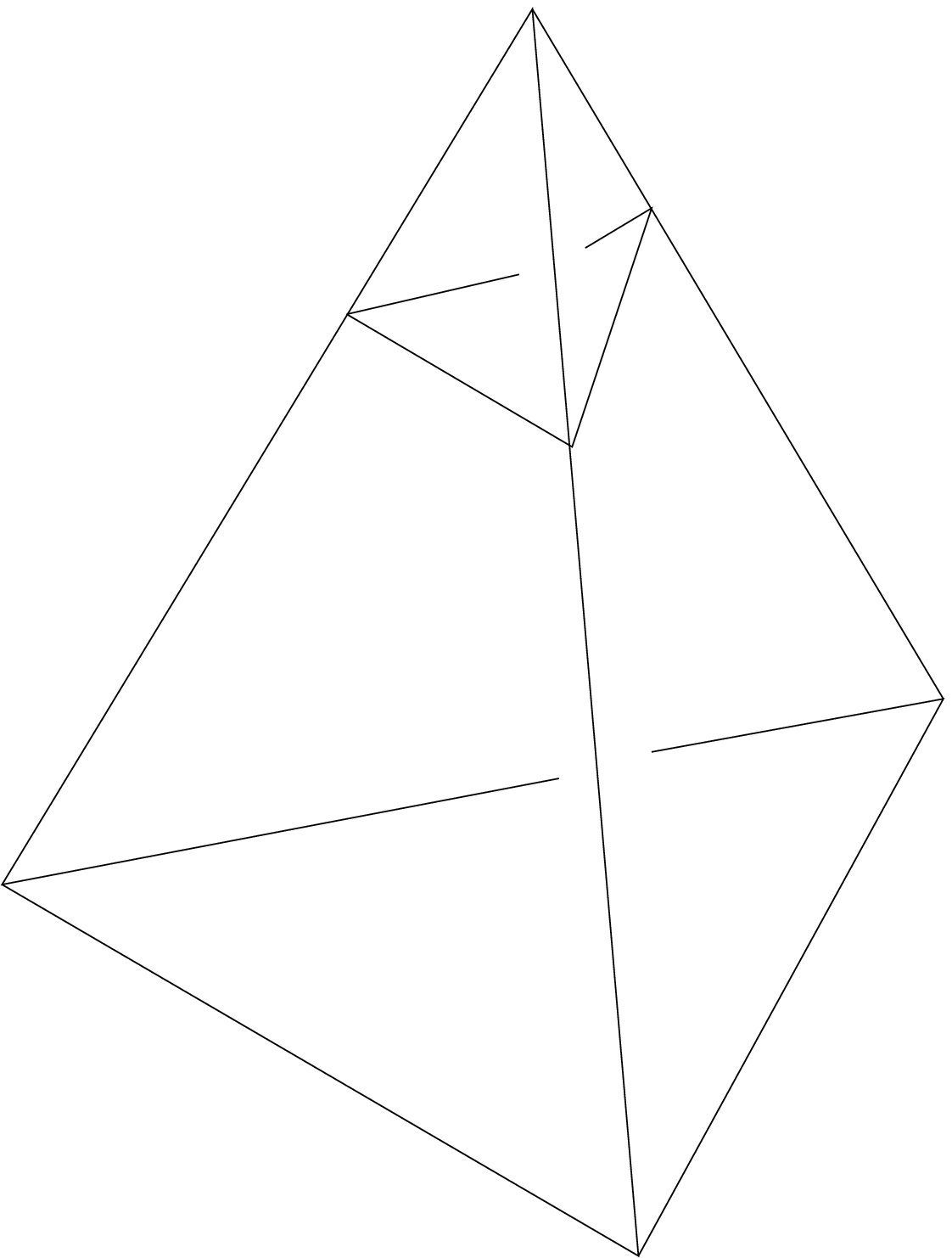, scale=0.1}}\ea+\ba{c}\mbox{\fig{file=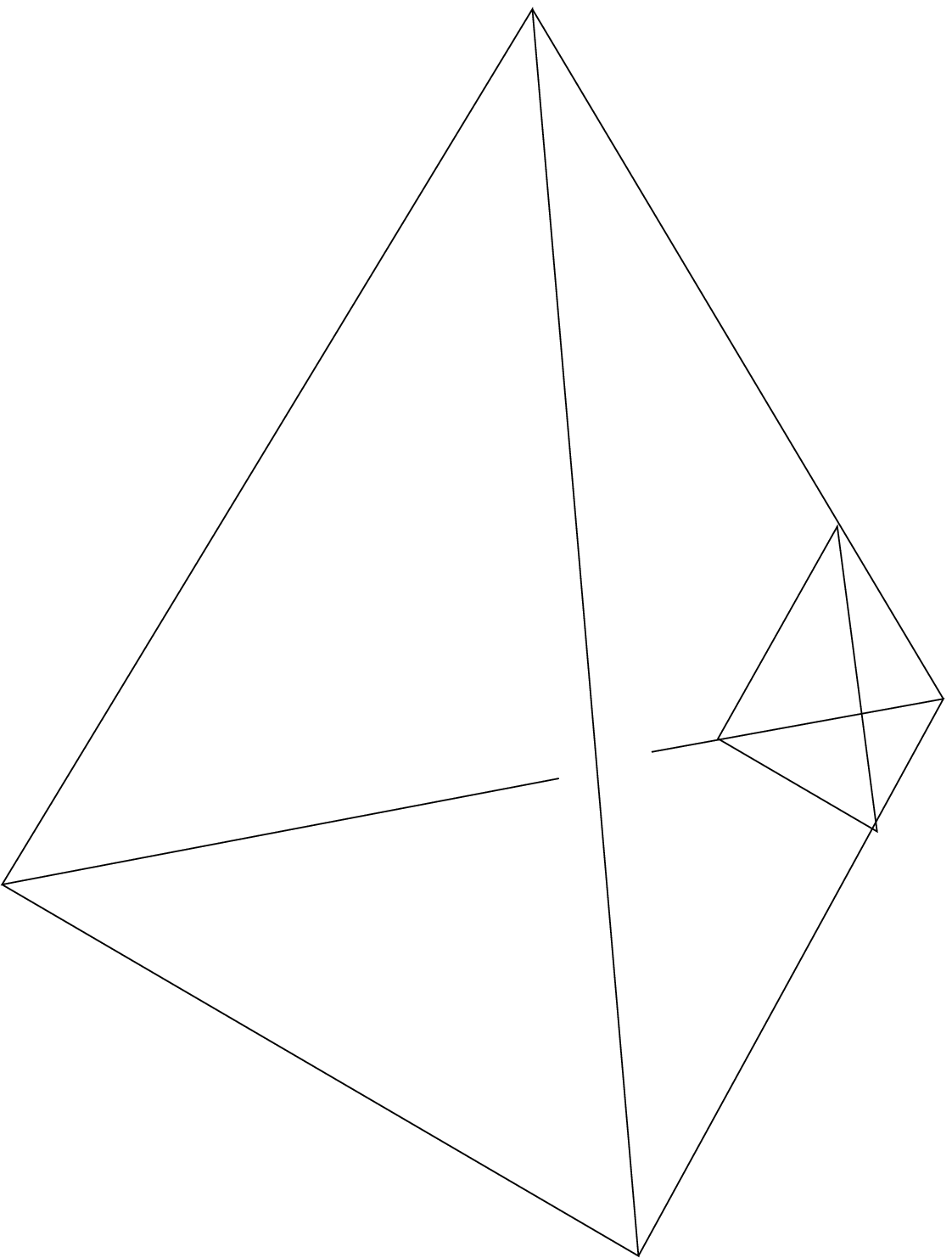, scale=0.1}}\ea+\ba{c}\mbox{\fig{file=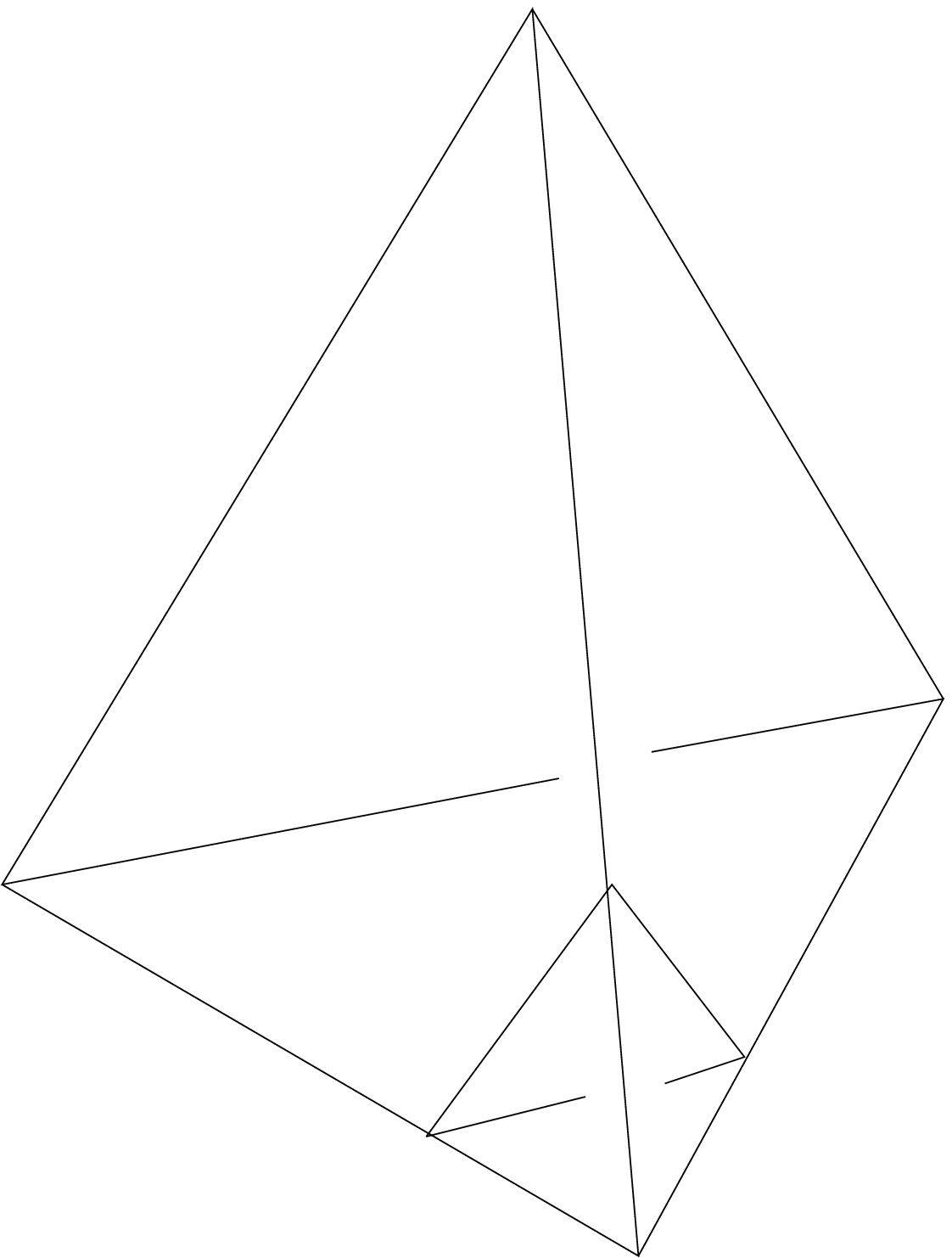, scale=0.1}}\ea\, ,\nonumber\\
c_3&=&B_+(6c_0^2c_1^2+4c_0^3c_2)=B_+(6c_1^2+4c_2)\nonumber\\
&=& B_+ \lp 6\ba{c}\mbox{\fig{file=tetraedru.eps, scale=0.07}}\ea\, \ba{c}\mbox{\fig{file=tetraedru.eps, scale=0.07}}\ea + 4\lp \ba{c}\mbox{\fig{file=t2-1.eps, scale=0.1}}\ea+\ba{c}\mbox{\fig{file=t2-2.eps, scale=0.1}}\ea+\ba{c}\mbox{\fig{file=t2-3.eps, scale=0.1}}\ea+\ba{c}\mbox{\fig{file=t2-4.eps, scale=0.1}}\ea\rp \rp.
\eeqa
We have left the last set of figures in the last equation above for the interested reader.

The general recursive solution for the equation \eqref{3DSE}, is given again by the Newton binomial formula:
\beqa
c_{n+1}=\sum_{k_1+\ldots+k_4=n}B_+(c_{k_1}\ldots c_{k_4}).
\eeqa

Identities \eqref{toate1} and \eqref{toate2} are also respected, the proof being analogous to the one of the previous section. Let us end this subsection by listing in the Table \ref{tabel2}  the polynomials $P_k^n$, 
which are 
 obtained analogously by applying the coproduct on the elements given in \eqref{frumusete3}:

\begin{table}[ht]
\caption{First values of the polynomial $P^n_k$ for the $3D$ case} 
\centering 
\begin{tabular}{c c c c c} 
\hline\hline 
$P^n_k$ & n=0 & n=1 &  n=2 & n=3 \\ [0.5ex] 
\hline 
k=0 & $1_\T$ & $c_1$ & $c_2$ & $c_3$ \\ 
k=1 &  & $1_\T$ & $4c_1$ & $6c_1^2+4c_2$ \\
k=2 &  &  &  $1_\T$ & $7c_1$\\
k=3 &  &  & & $1_\T$\\
[1ex] 
\hline 
\end{tabular}
\label{tabel2} 
\end{table}

\subsection{The $4D$ case}

\begin{figure}
\centerline{\epsfig{figure=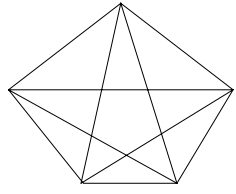,width=3cm} }
\caption{A $4-$simplex is the building block of the core Hopf algebra of  $4D$ SFs. It plays the same role as the triangle in the $2D$ construction or the tetrahedron in $3D$, being the graduation one generator in the algebra.}
\label{hyper}
\end{figure}

The $4D$ case is treated along the same lines. The building brick is now the $4-$simplex of Fig. \ref{hyper}.

The appropriate equation to investigate (generalizing \eqref{3DSE}) is
\beqa
\label{4DSE}
X=\unu+tB_+ (X^5),
\eeqa
since the maximal number of insertion places is now five.

The solution of this equation (in the first four orders of the development in the constant $t$) is:
\beqa
\label{frumusete4}
c_0&=&\unu,\nonumber\\
c_1&=&B_+(c_0^5)=B_+(\unu)=
\ba{c}\mbox{\fig{file=hyper.eps, scale=0.07}}\ea
\, ,\nonumber\\
c_2&=&5B_+(c_0^4c_1)=5B_+(c_1)
=4B_+\left(\ba{c}\mbox{\fig{file=hyper.eps, scale=0.07}}\ea\right),
\nonumber\\
c_3&=&B_+(10c_0^3c_1^2+5c_0^4c_2)=B_+(10c_1^2+5c_2).
\eeqa

The general solution writes
\beqa
c_{n+1}=\sum_{k_1+\ldots+k_5=n}B_+(c_{k_1}\ldots c_{k_5}).
\eeqa

As in the previous cases, identities \eqref{toate1} and \eqref{toate2} hold in the same manner. Let us list here the set of the first polynomials $P_k^n$:

\begin{table}[ht]
\caption{First values of the polynomial $P^n_k$ for the $3D$ case} 
\centering 
\begin{tabular}{c c c c c} 
\hline\hline 
$P^n_k$ & n=0 & n=1 &  n=2 & n=3 \\ [0.5ex] 
\hline 
k=0 & $1_\T$ & $c_1$ & $c_2$ & $c_3$ \\ 
k=1 &  & $1_\T$ & $5c_1$ & $10c_1^2+5c_2$ \\
k=2 &  &  &  $1_\T$ & $9c_1$\\
k=3 &  &  & & $1_\T$\\
[1ex] 
\hline 
\end{tabular}
\label{tabel3} 
\end{table}

\bigskip

We end this section with the following comparison. 
In the rooted tree Hopf algebra literature, general combinatorial Dyson-Schwinger equations can be considered (see for example \cite{teza-yeats}).
Nevertheless, let us note that, as already announced in the previous section, combinatorial Dyson-Schwinger equations of the following particular form are generally used (see for example \cite{bk, anatomy, joc}):
\beqa
\label{DSE-k}
X=1_{\mathbb T}+\sum_{n=1}^\infty t^n \omega_n B_+^{d_n} (X^{n+1}),
\eeqa
where $\omega_n$ are scalars and $(B_+^{d_n})$ is a collection of Hochschild one-cocycles on the algebra (see again \cite{bk} for details). The main difference with the equation \eqref{DSE} (or \eqref{3DSE} or \eqref{4DSE} that we use here) is in the power of the constant $t$. This leads to crucial differences in the calculus of the polynomials $P^n_k$. For example, when considering the equation
\beqa
X=1_{\mathbb T}+ t^2 \omega_n B_+ (X^3),
\eeqa
one  obtains
\beqa
c_0&=&\unu,\nonumber\\
c_1&=&0,\nonumber\\
c_2&=&B_+(c_0)=B_+(\unu),\nonumber\\
c_3&=&3B_+(c_0c_1)=0,\nonumber\\
c_4&=&3B_+(c_0c_1^2+c_0^2 c_2)=3B_+(c_2)
\eeqa
and so on. One can directly see that this is different from equation \eqref{frumusete} (or \eqref{frumusete3} or \eqref{frumusete4}). This further leads to a different set of polynomials that the ones listed in Table \ref{tabel1}. To end this discussion, let us also remark, that the polynomials $P_k^n$ associated with the combinatorial Dyson-Schwinger equation \eqref{DSE-k} do {\it not} depend on the scalars $\omega_n$ or on $B_+^{d_n}$. This is not the case for the polynomials exhibited in this paper, which are different for the $2$, $3$ or $4D$ cases (see for example Tables \ref{tabel1}, \ref{tabel2} and \ref{tabel3}).

\section{Comments on the physical relevance of the approach; example}
\renewcommand{\theequation}{\thesection.\arabic{equation}}    
\setcounter{equation}{0}

As already stated in the Introduction, a general power counting theorem for SF models is not known today; in \cite{fm}, an algebraic structure was introduced where the coproduct $\Delta_M$ sums over all sub-SFs. In the Hopf algebra defined in this paper, the same definition of the coproduct is kept, {\it i. e.} one sums over all sub-SFs. These constructions, both the one in \cite{fm} and the one here can be seen as a first attempt towards better understanding the renormalizability properties of SF models.

Furthermore, we have also argued above that this type of algebraic structure can be well-suited to deal with quantum gravity because of the following argument. Hopf primitives ({\it i. e.} the elements of the Hopf algebra which have a trivial coproduct) of perturbative quantum gravity are one-loop graphs. Hopf primitives are directly related to the primitive divergent graph of a field theory. Therefore, it appears natural, from this point of view, to consider core Hopf algebra ({\it i. e.} Hopf algebra in which the coproduct sums on all respective sub-SFs) as an interesting structure to investigate.

\begin{figure}
\centerline{\psfig{figure=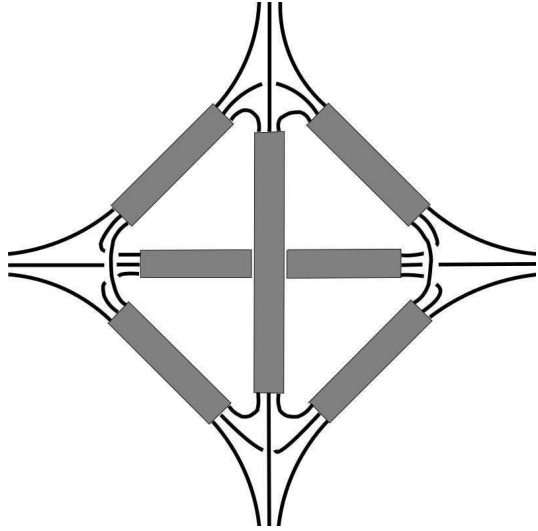,width=7cm} }
\caption{$3D$ group field theoretical graph corresponding to the SF of Fig. \ref{3dsf}. This graph contains one bubble and its Feynman amplitude is divergent.}
\label{gft}
\end{figure}

Finally, let me give one additional argument, using this time the group field theoretical approach. We focus on the $3D$ case (the $2D$ one being trivial). One can associate the SF of Fig. \ref{3dsf} - the divergent quantity and also the Hopf primitive here - to the graph of Fig. \ref{gft}. One can easily identify a bubble (a closed $3-$dimensional region of the graph or a closed bi-circuit) in this graph. This topological notion of bubble (see for example \cite{fgo}) is the natural generalization of the notion of face (closed circuits in the graph). Let us also emphasize that in \cite{fgo} an algorithm for identifying the bubbles of a generic $3-$dimensional group field theory was given.

The Feynman amplitude associated with the graph of Fig. \ref{gft} is divergent \cite{com}. This fact is thus a further indication for choosing the coproduct used here and in \cite{fm}.

\medskip

Let us now comment further on the significance of the combinatorial Dyson-Schwinger equation \eqref{DSE} (and its generalization to higher dimensions). As already mentioned above, this is the analogue of a cubic combinatorial Dyson-Schwinger equation from the QFT framework. The field action related to this equation is the group field theoretical one, which in the $3D$ case writes 
\beqa
S[\phi]&=&\frac 12 \int d g_1 d g_2 d g_3 \phi (g_1,g_2,g_3)\phi(g_3,g_2,g_1),\\
&+&\frac{\lambda}{4}\int \phi(g_1,g_2,g_3)\phi(g_3,g_4,g_5)\phi(g_5,g_6,g_1)\phi (g_6,g_4,g_2).\nonumber
\eeqa
The integrations over the group (left here implicit in the interaction term) are performed as usually with the invariant de Haar mesure.

This equation is the combinatorial backbone of non-perturbative QFT. The analytic Dyson-Schwinger equation (the one used in physics) is obtained by applying the renormalized Feynman rules to the combinatorial one. Let us recall here that Dyson-Schwinger equations are quantum equations of motion for the Green (or Schwinger) functions, being thus a crucial tool of any QFT. We argue that it is thusly justified to analyse (here from a combinatorial point of view) such an equation in our efforts towards a better understanding of a quantum formulation of gravity.

As we will also comment on in the following section, it would be interesting to adapt such tools to the group field theoretical approach also (which is naturally suited for such a study) and then to compare the results with the one obtained in this paper.

\section{Conclusions and perspectives}
\renewcommand{\theequation}{\thesection.\arabic{equation}}    
\setcounter{equation}{0}

We have thus explicitly exhibited, in the framework of the SF formalism, some combinatorial notions which  naturally appear in QFT. The $2$, $3$ and $4D$ cases have been analyzed and some non-trivial examples have been worked out as an illustration of our results.

The correspondence of the Hopf algebra $\T$ defined here is done with the Hopf algebra of rooted trees and not with the Connes-Kreimer Hopf algebra of Feynman graphs. This comes  from the fact that we don't deal with overlapping SFs. In QFT, dealing with overlapping divergences by rooted trees is also more involved (see \cite{over1, over2}).
A $1PI$ Feynman graph can be uniquely represented by a rooted tree (with labels on each vertex corresponding to the associated subgraph) iff all subdivergences are nested and not overlapping and if there is only one way to make each insertion. It would be interesting to investigate whether or not a correspondence between overlapping SFs and the Connes-Kreimer algebra of Feynman graph can be obtained.

Nevertheless, let us stress on the following issue. As already mentioned in the Introduction, in commutative (resp. noncommutative) QFT, behind the combinatorial properties investigated here lies the physical principle of locality (or resp. ``Moyality''). The renormalizability of local theories (or of non-local scalars models on the Moyal space - see \cite{ncqft}) is by now well understood. One cannot say today that this is also the case for quantum gravity models. It appears to us of crucial importance to investigate whether or not a generalization of the principles of locality (or ``Moyality'') can exist. This new type of principle could be related, from a combinatorial point of view, to the fact that the triangular character of SFs reproduces itself when inserting SFs into SFs, having thus some  kind of ``triangularity'' 
(or similarly {``simplexality''} for higher dimensions). 

A promising way of approaching the renormalizability of quantum gravity can be a thorough study of group field theoretical models (see for example \cite{gft}). These models 
were developed as a generalization of $2D$ matrix models to $3D$ or $4D$. Thus, group field theoretical models are duals to the Ponzano-Regge model, when considering the $3D$ gravity, or to the Ouguri model, when considering the $4D$ one.

These models can be seen nowadays not only as a technical tool but as a proposition for a quantum formulation of gravitation. Behind this lies the idea that group field theories are theories of space-time, while
QFT are theories on space-time. Feynman graphs of these models are tensor graphs, a natural generalization of the matrix graphs of noncommutative QFT. Recently, 
insights on the renormalizability of $3D$ models have been given 
\cite{fgo, ultim}.

A perspective to be mentioned here is  the investigation of the combinatorial properties studied in this paper within this new context of group field theory. Moreover, a comparison of the results 
 obtained from this program
with the results of this paper could offer a better understanding of the physical properties of these gravitational models.

\bigskip

\noi
{\bf Acknowledgment:} 
D. Kreimer, K. Noui, K. Yeats and M. Smerlak are acknowledged for discussions. 
The author  was partially supported by the CNCSIS grant ``Idei'' 454/2009, ID-44.


\begin{thebibliography}{99}

\bibitem{book-connes}
``Noncommutative Geometry''
Alain Connes, 
Academic Press, San Diego, 1994.

\bibitem{dfr}
S.~Doplicher, K.~Fredenhagen and J.~E.~Roberts,
  ``Space-time quantization induced by classical gravity,''
  Phys.\ Lett.\  B {\bf 331}, 39 (1994).

\bibitem{book}
``Quantum gravity'', Carlo Rovelli, Cambridge University Press, 2004.

\bibitem{gft}
 L.~Freidel,
 ``Group field theory: An overview,''
  Int.\ J.\ Theor.\ Phys.\  {\bf 44}, 1769 (2005)
  [arXiv:hep-th/0505016].
 D.~Oriti,
  ``Quantum gravity as a group field theory: A sketch,''
  J.\ Phys.\ Conf.\ Ser.\  {\bf 33}, 271 (2006)
  [arXiv:gr-qc/0512048].
 D.~Oriti,
``The group field theory approach to quantum gravity,''
  arXiv:gr-qc/0607032.



\bibitem{fm}
F.~Markopoulou,
  ``Coarse graining in spin foam models,''
  Class.\ Quant.\ Grav.\  {\bf 20}, 777 (2003)
  [arXiv:gr-qc/0203036].

\bibitem{fm-primu}
F.~Markopoulou,
``An algebraic approach to coarse graining,''
  arXiv:hep-th/0006199.

\bibitem{ck}
  A.~Connes and D.~Kreimer,
  ``Renormalization in quantum field theory and the Riemann-Hilbert  problem.
  I: The Hopf algebra structure of graphs and the main theorem,''
  Commun.\ Math.\ Phys.\  {\bf 210}, 249 (2000)
  [arXiv:hep-th/9912092].


\bibitem{fab}
 A.~Tanasa and F.~Vignes-Tourneret,
  ``Hopf algebra of non-commutative field theory,''
J.\ Noncomm.\ Geom. {\bf 2}, 125-139  (2008).
arXiv:0707.4143[math-ph].

\bibitem{io-kreimer}
A. Tanasa and D. Kreimer, ``Combinatorial Dyson-Schwinger equation in noncommutative field theory'', submitted to { J. Noncomm. Geom.},  arXiv:0907.2182 [hep-th].

\bibitem{book2}
``Form perturbative to constructive renormalization'', Vincent Rivasseau, Princeton University Press, 1992.

\bibitem{rev}
V.~Rivasseau,
  ``Non-commutative renormalization,''
  arXiv:0705.0705 [hep-th].

\bibitem{beta-GMRT}
  J.~B.~Geloun and A.~Tanasa,
  ``One-loop $\beta$ functions of a translation-invariant renormalizable
  noncommutative scalar model,'' Lett. Math. Phys. {\bf 86}, 19, (2008).
  arXiv:0806.3886 [math-ph].

\bibitem{core}
  S.~Bloch and D.~Kreimer,
  ``Mixed Hodge Structures and Renormalization in Physics,''
  arXiv:0804.4399 [hep-th].

\bibitem{k-qg}
D. Kreimer, ``A remark on quantum gravity'', Annals Phys. {\bf 323}, 49 (2008), arXiv:0705.3897[hep-th].



\bibitem{core-suite}
D. Kreimer, ``The core Hopf algebra'', arXiv:0902.1223 [hep-th]; 
D.~Kreimer and W.~D.~van Suijlekom,
  ``Recursive relations in the core Hopf algebra,''
  arXiv:0903.2849 [hep-th].



\bibitem{bk}
  C.~Bergbauer and D.~Kreimer,
  ``Hopf algebras in renormalization theory: Locality and Dyson-Schwinger
  equations from Hochschild cohomology,''
  IRMA Lect.\ Math.\ Theor.\ Phys.\  {\bf 10}, 133 (2006)
  [arXiv:hep-th/0506190].

\bibitem{yeats}
D.~Kreimer and K.~Yeats,
 ``Recursion and growth estimates in renormalizable quantum field theory,''
  Commun.\ Math.\ Phys.\  {\bf 279}, 401 (2008)
  [arXiv:hep-th/0612179].



\bibitem{sf}
D.~Oriti,
 ``Spacetime geometry from algebra: Spin foam models for non-perturbative
 quantum gravity,''
  Rept.\ Prog.\ Phys.\  {\bf 64}, 1489 (2001)
  [arXiv:gr-qc/0106091].

\bibitem{eprl}
  J.~Engle, R.~Pereira and C.~Rovelli,
  ``The loop-quantum-gravity vertex-amplitude,''
  Phys.\ Rev.\ Lett.\  {\bf 99}, 161301 (2007)
  [arXiv:0705.2388 [gr-qc]].

\bibitem{fk}
  L.~Freidel and K.~Krasnov,
  ``A New Spin Foam Model for 4d Gravity,''
  Class.\ Quant.\ Grav.\  {\bf 25}, 125018 (2008)
  [arXiv:0708.1595 [gr-qc]].

\bibitem{rt}
D.~Kreimer,
  ``On the Hopf algebra structure of perturbative quantum field theories,''
  Adv.\ Theor.\ Math.\ Phys.\  {\bf 2}, 303 (1998)
  [arXiv:q-alg/9707029].

\bibitem{rt-ck}
A.~Connes and D.~Kreimer,
  ``Hopf algebras, renormalization and noncommutative geometry,''
  Commun.\ Math.\ Phys.\  {\bf 199}, 203 (1998)
  [arXiv:hep-th/9808042].



\bibitem{foissy1}
L. Foissy, ``Les alg\`ebres de Hopf des arbres enracin\'es'' I, Bull. Sci. Math. {\bf 126}, 193 (2002) (in French).

\bibitem{foissy2} L. Foissy, ``Les alg\`ebres de Hopf des arbres enracin\'es'' II, Bull. Sci. Math. {\bf 126}, 249 (2002) (in French).


\bibitem{exact}
  D.~J.~Broadhurst and D.~Kreimer,
  ``Exact solutions of Dyson-Schwinger equations for iterated one-loop
  integrals and propagator-coupling duality,''
  Nucl.\ Phys.\  B {\bf 600}, 403 (2001)
  [arXiv:hep-th/0012146].

\bibitem{teza-yeats}
K.~A.~Yeats,
  ``Growth estimates for Dyson-Schwinger equations,'' PhD thesis, 
  arXiv:0810.2249 [math-ph].

\bibitem{anatomy}
  D.~Kreimer,
  ``Anatomy of a gauge theory,''
  Annals Phys.\  {\bf 321}, 2757 (2006)
  [arXiv:hep-th/0509135].

\bibitem{joc}
  D.~Kreimer,
  ``Factorization in quantum field theory: An exercise in Hopf algebras and
  local singularities,'' Contributed to Les Houches School of Physics: Frontiers in Number Theory, Physics and Geometry, Les Houches, France, 9-21 Mar 2003,
  arXiv:hep-th/0306020.

\bibitem{fgo}
 L. Freidel, R. Gurau and D. Oriti, ``Group field theory renormalization - the 3d case: power counting of divergences'', Phys. Rev. D {\bf 87}, 044007 (2009).

\bibitem{com}
J. Ben Geloun, T. Krajewski, J. Magnen, V. Rivasseau, in progress, M. Smerlak, in progress.

\bibitem{over1}
D.~Kreimer,
``On overlapping divergences,''
  Commun.\ Math.\ Phys.\  {\bf 204}, 669 (1999)
  [arXiv:hep-th/9810022].

\bibitem{over2}
H.~Figueroa and J.~M.~Gracia-Bondia,
``Combinatorial Hopf algebras in quantum field theory. I,''
  Rev.\ Math.\ Phys.\  {\bf 17}, 881 (2005)
  [arXiv:hep-th/0408145].



\bibitem{ncqft}
H. Grosse and R. Wulkenhaar, Renormalization of $\phi^4$-theory on noncommutative $\RR^4$ in the matrix base'', Commun. Math. Phys. {bg 256}, 305 (2005);
R.~Gurau, J.~Magnen, V.~Rivasseau and A.~Tanasa,
  ``A translation-invariant renormalizable non-commutative scalar model,''
  arXiv:0802.0791 [math-ph], Commun. Math. Phys. {\bf 287}, 275, (2008). 

\bibitem{ultim}
J. Magnen, K. Noui, V. Rivasseau and M. Smerlak, ``Scaling behaviour of three-dimensional group field theory'', arXiv:0906.5477, Class. Quant. Grav. (in press);
R.~Gurau,
``Colored Group Field Theory,''
  arXiv:0907.2582 [hep-th]; R.~Gurau,
  ``Topological Graph Polynomials in Colored Group Field Theory,''
  arXiv:0911.1945; J.~B.~Geloun, J.~Magnen and V.~Rivasseau,
  ``Bosonic Colored Group Field Theory,''
  arXiv:0911.1719 


\end{thebibliography}
\end{document}